# THE GOLDMAN-HODGKINS-KATZ EQUATION, REVERSE-ELECTRODIALYSIS, AND EVERYTHING IN BETWEEN

Yoav Green[*]

Department of Mechanical Engineering, Ben-Gurion University of the Negev, Beer-Sheva 8410501, Israel

**Abstract.**

In the past eighty years, the Goldman-Hodgkins-Katz (GHK) equation has been the gold-standard framework for interpreting countless biological and physiological experiments and simulations that involve ion transport in nanopores/nanochannels/ion-channels subjected to a combined ionic concentration and electric potential gradients. In this work, we revisit the mathematical derivation used to develop the GHK model and show that this model is internally inconsistent. In particular, we show that its infamous assumption of a constant electric field is incorrect, which leads to substantial errors, including the inability of this model to satisfy local and global electroneutrality. Then, leveraging key insights from the field of reverse electrodialysis (RED), we derive a new internally consistent model that does not assume that the electric field is constant and satisfies electroneutrality. This new model has several advantages. First, while the mathematics are substantially more complicated, the derivation doesn't include ad-hoc assumptions, and the model is internally consistent. Second, the new solution connects the two realms of GHK and RED, which consider the same equations but in opposing limits, negligible or substantial surface charge density effects, respectively. Third, while the expressions for the new model are complicated, the new model can be reduced to several limits, which allows for a much easier and more straightforward analysis. Finally, all of our newly derived results show remarkable correspondence to non-approximated numerical simulations. This work provides a brand-new framework for interpreting (and reinterpreting) ion transport experiments in any charge-selective system.

[*] Email: yoavgreen@bgu.ac.il
ORCID number:: 0000-0002-0809-6575

## I. INTRODUCTION

Ion transport through charged nanochannels, nanopores, and ion channels is ubiquitous in technology (desalination [1–5], energy harvesting [6–9], fluid-based circuits [10–14] (including memristors [15–17]), and biosensing [18–21]), DNA sequencing [20,22–26], and nature (basic physiological phenomena [9,27–32]). Importantly, regardless of the system, the application, or the purpose, ion transport in all these systems is governed by the same equations and shares the same underlying physics principles. Thus, knowledge of one system inexplicitly yields knowledge of the remaining systems.

We recently developed a new mathematical method to investigate and characterize reverse-electrodialysis (RED) systems (energy harvesting via concentration gradients) [33]. As pointed to us by Prof. Slaven Garaj, this new model also has implications for the classical Goldman-Hodgkins-Katz (GHK) theory. At his encouragement, we have used this new method to revisit the GHK theory used for the investigation of ion transport across biological membranes, with the intention of seeing whether these two models overlap or not and, if not, why.

The GHK theory was derived approximately eight decades ago in the two seminal works of Goldman [27] and Hodgkins and Katz [28], and since then, it has served as the cornerstone theory of the ion-channel community. The theory has been supposedly verified by countless experiments and, thus, is supposedly beyond reapproach. However, in the course of revisiting this classical theory, we found that the assertion that the electric field is uniform, and all quantities calculated using this assertion, leads to several problems and inconsistencies.

Even though the number of papers citing GHK is extremely vast, in this work, we will not be able, nor will we try to address the question as to why, thus far, these inconsistencies were not uncovered. Instead, we will focus on two much easier problems: 1) presenting the inconsistencies and their outcomes and 2) deriving an alternative solution for two ionic species that connect (a modified-)GHK and RED in an internally consistent manner.

Our paper is structured as follows:
- In Section II, we will demonstrate/establish that the equations and boundary conditions governing GHK and RED are identical – these are the Poisson-Nernst-Planck equations subject to a concentration gradient and an electric potential gradient. The difference between GHK and RED is that GHK is derived for the cases where the excess counterion concentration is zero ($\tilde{\Sigma}_s = 0$) – this parameter will be defined later more accurately – while RED systems operate in the opposite limit of very large $\tilde{\Sigma}_s$. In this work, we will derive a solution for arbitrary $\tilde{\Sigma}_s$. We will show that while the RED limit is correct, the GHK limit requires modification. In this section, we also present the key characteristics of the current-voltage, $\tilde{i} - \tilde{V}$, response and discuss them with regard to GHK and RED.

- Before suggesting modifications to GHK, it is essential to understand the model and its embedded assumptions. Thus, before deriving the new general solution, in Sec. III, we rederive the GHK model, step-by-step, following Goldman's original work [27] (for the sake of simplicity and brevity, we will primarily focus on Goldman's work and only briefly discuss the work of Hodgkins and Katz [28], which utilizes the same assumptions as Goldman's). In this section, we will also present non-approximated numerical simulations that disagree with Goldman's results. To explain the disagreements, we will discuss how the various assumptions affect the derived quantities.

- In Sec. IV, we will derive a new general solution, in non-dimensional formulation, that holds for arbitrary $\tilde{\Sigma}_s$. From this theory, we will derive expressions for all the main transport characteristics.

- Section V has three purposes. First, we will present the dimensional quantities of all the main transport characteristics. Second, we will demonstrate that, in various limits, the complicated expressions for the general theory can be reduced to three simpler models. The final purpose will be to analyze each of these simple models (from simplest to complicated) to build the required framework with which the general solution can be analyzed. Throughout, we will show that our new theory is substantiated by non-approximated numerical simulations.

- In Sec. VI, we finish with short concluding remarks.

## II. BACKGROUND

### A. System overview

Figure 1 presents a schematic of a prototypical system in RED as well as in GHK, where the nano-/ion- channel is embedded in an isolating material such that the fluxes can go only through the channel. The system is subjected to a combined voltage drop, $\tilde{V}$, and concentration gradient due to asymmetric bulk concentrations denoted by $\tilde{c}_{\text{left}}$ and $\tilde{c}_{\text{right}}$, creating an electrical current density, $\tilde{i}$.

In this work, we will assume uniform properties across the the cross-section such that all extensive properties depend on the flux densities multiplied by the cross-section area, $\tilde{A}$. For example, the electrical current density is related to the electric current, $\tilde{I}$, by $\tilde{i} = \tilde{I} / \tilde{A}$. In the remainder, we will refer to the current and current density interchangeably. Since the 3D Ohmic conductance

$$\tilde{G}_{\text{Ohmic}} = \tilde{I}/\tilde{V} = \tilde{g}_{\text{Ohmic}}\tilde{A}, \qquad (1)$$

depends on the area, we will use the conductance-density

$$\tilde{g}_{\text{Ohmic}} = \tilde{I}/(A\tilde{V}) = \tilde{i}/\tilde{V}, \qquad (2)$$

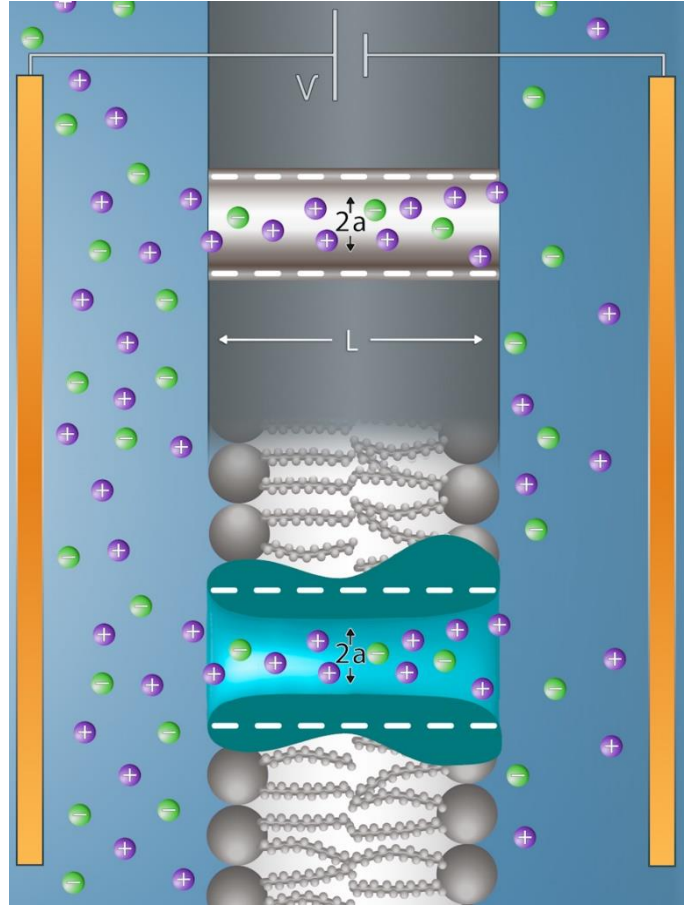

**Figure 1**. Schematic of two separate nanofluidic systems that have been merged into one. On the top, we have a nanopore in a simple dielectric material, while on the bottom, we have an ion channel embedded into biological membranes. Both (or either) channels are subjected to a combined voltage drop $\tilde{V}$ (defined positive from left to right) and asymmetric concentrations by $\tilde{c}_{\text{left}}$ and $\tilde{c}_{\text{right}}$ at the two respective ends of the nanochannel or ion channel. In general $\tilde{c}_{\text{left}}$ and $\tilde{c}_{\text{right}}$ are not equal. Here, we have depicted that $\tilde{c}_{\text{left}} > \tilde{c}_{\text{right}}$, however, the reverse situation is also allowable. Both channels are charged with an embedded surface charge density (here, negative and denoted by white minus signs). The surface charge leads to a surplus of purple counterions (here, positive) over the green coions (here, negative). The dielectric material or biological membrane are completely isolating such that fluxes are allowed only through the channel. The channel length is $\tilde{L}$. The characteristic lateral length of the systems is the (average-) radius of the pore is $\tilde{a}$.

which is area-independent. In the remainder, to avoid the need to multiply by the area, our focus will be on the 1D quantities, such as $\tilde{g}_{\text{Ohmic}}$. Also, for the sake of simplicity, we will continue to refer to the Ohmic conductance density as the Ohmic conductance or simply as the "conductance".

Finally, note that in Figure 1, we have portrayed the geometry of the nanopore/nanochannel as circular. However, this assumption can be relaxed so long as $\tilde{L} \gg \tilde{A}^{1/2}$. Then, our results will hold for nanochannels of arbitrary cross-sections (parallel plates, ellipses, and more). In reality, this would also hold for ion channels [which are typically very cylindrical(-like) but do not have an axisymmetric axis].

### B. The current-voltage response

These systems are characterized by their $\tilde{i} - \tilde{V}$ curves. It has long been known that if the concentrations at the two ends of the channel are symmetric, the $\tilde{i} - \tilde{V}$ goes through the origin (blue-line in Figure 2), whereas if the concentrations at the two ends of the system are asymmetric, the $\tilde{i} - \tilde{V}$ is shifted from the origin (red-line in Figure 2), yielding two important intercept points: the zero-current voltage, $\tilde{V}_{\tilde{i}=0}$, and zero-voltage current, $\tilde{i}_{\tilde{V}=0}$.

It is perhaps surprising that the emphasis and interests of RED and GHK on $\tilde{V}_{\tilde{i}=0}$ and $\tilde{i}_{\tilde{V}=0}$ is rather complementary. In the field of RED, the zero-voltage current, $\tilde{i}_{\tilde{V}=0}$, is the desired quantity of interest as it is a proxy for the power density that can be harvested. In biology, the zero-current voltage, $\tilde{V}_{\tilde{i}=0}$, is the quantity of interest as it serves as a proxy for the electric potential barrier that needs to be overcome for physiological events to occur. As we will show in this work, in a complete and internally consistent theory, knowledge of one quantity inexplicitly leads to knowledge of the other.

Perhaps the most important (electrical) characterizer of nanofluidic systems is the low-voltage-low current Ohmic conductance of the system given by $\tilde{g}_{\text{Ohmic}} = \tilde{i}/\tilde{V}$. In this work, we will also consider the conductance at the two intercept points $\tilde{V}_{\tilde{i}=0}$ and $\tilde{i}_{\tilde{V}=0}$ (Figure 2). Thus, it is important to make a distinction between the two. The well-investigated Ohmic conductance corresponds to the slope at $\tilde{V}_{\tilde{i}=0}$, such that $\tilde{g}_{\text{Ohmic}} = \tilde{g}_{\tilde{i}=0}$ [the conductance can be calculated either using the linear approximation $\tilde{i}/\tilde{V}$ or $(d\tilde{i}/d\tilde{V})_{\tilde{i}=0}$]. It can already be noted that the slope of conductance at $\tilde{i}_{\tilde{V}=0}$ is different than that at $\tilde{V}_{\tilde{i}=0}$ such that $\tilde{g}_{\text{Ohmic}} \neq \tilde{g}_{\tilde{V}=0}$ (this will be proved later in Sec. V.C). As we will later show, the differential slope, i.e., the differential conductance, $\tilde{g}_{\tilde{V}=0} = (d\tilde{i}/d\tilde{V})_{\tilde{V}=0}$ needs to be numerically calculated through a transcendental equation (this is discussed later).

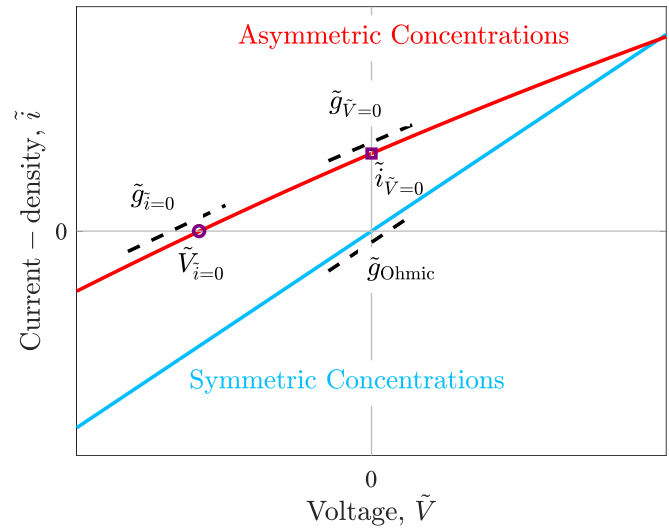

**Figure 2**. Schematics of the current(-density)-voltage $\tilde{i}-\tilde{V}$ responses of nanofluidic subject to symmetric concentrations (blue) and asymmetric concentrations (red). When the concentrations are symmetric, the response is symmetric, linear, and crosses through the origin. When the concentrations are asymmetric, the $\tilde{i}-\tilde{V}$ is shifted relative to the origin and is characterized by the two intercepts: zero-current voltage, $\tilde{V}_{\tilde{i}=0}$, and zero-voltage current, $\tilde{i}_{\tilde{V}=0}$. In general, the response is not purely linear, whereby the slopes at $\tilde{i}=0$ and $\tilde{V}=0$ are not the same [i.e., $\tilde{g}_{\text{Ohmic}} = \tilde{g}_{\tilde{i}=0} = (d\tilde{i}/d\tilde{V})_{\tilde{i}=0} \neq (d\tilde{i}/d\tilde{V})_{\tilde{V}=0} = \tilde{g}_{\tilde{V}=0}$].

Note that the $\tilde{i}-\tilde{V}$ shown in Figure 2 for the symmetric concentration appears to be linear linear always. This is because we have not accounted for the effects of the adjoining reservoirs. If they are accounted for, the additional resistances of the channels themselves and the access resistances become non-negligible. At sufficiently large enough voltages, the $\tilde{i}-\tilde{V}$ deviates from the linear relation. Thereafter, limiting [34–37] and over-limiting currents [38–44] will appear. In this work, we will ignore the effects of these larger channels and focus only on the nanochannel.

1. **Ohmic conductance**

Since the pioneering work of Stein et al. [45], it has been known that the conductance behaves in the following manner (a mathematical equation will be given later): at very high concentrations, when the effects of the surface charge are negligible, the conductance is linear with the bulk concentration, while at low concentrations, the conductance saturates to a value that depends on the surface charge density (Figure 3). The difference between the two states is that at low concentrations (Figure 3 left inset), the surface charge density is responsible for completely excluding ions of the same sign (coions) and allowing ions of the opposite sign (counterion) to go through freely. Naturally, since this state is termed 'highly selective' or 'ideally selective' – ideally selective systems will later be denoted as $\Sigma_s = \tilde{\Sigma}_s / \tilde{c}_{bulk} \gg 1$. In contrast, in the former state of high concentrations (Figure 3 right inset), the effects of the surface charge are so negligible such that both coions and counterions are transported in an equal manner, and there is no selectivity. Naturally, this state is termed as 'non-selective' or 'vanishingly selective'. Vanishingly selective systems will later be denoted as $\Sigma_s \ll 1$.

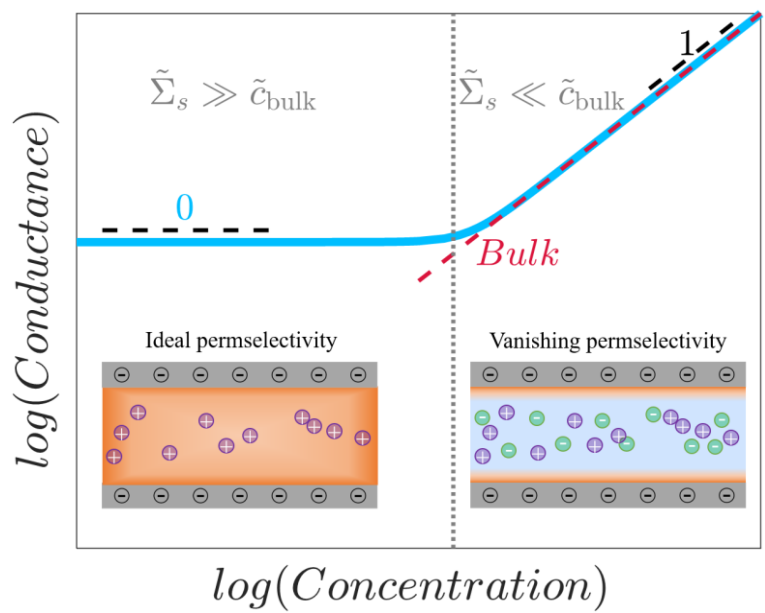

**Figure 3**. Schematic of the Ohmic conductance vs the bulk concentration. At high concentrations, the conductance is linear in the concentration, while at low concentrations, the conductance is concentration-independent. The transition from one limit to the other depends on the ratio of the excess counterion concentration, $\tilde{\Sigma}_s$, to the bulk concentration $\tilde{c}_{bulk}$. (Bottom-right inset)At high concentrations, the effects of the surface charge are limited to the thin electric double layers (denoted by the not-to-scale- orange regions), such that most of the current is transferred through the bulk (denoted by blue regions), such that the system is vanishingly selective. (Bottom-left inset) In contrast, at low concentrations, the electric double layers span the entire channel, such that the surface charge completely expels the coions, and the channel is ideally selective.

Another interesting difference between the two communities (ion-channels/GHK and nanochannels/RED) is the embedded assumptions used in the modeling. In order to harvest an electrical current, some symmetry-breaking between the counterion and coions is needed. Thus, RED systems typically operate with some degree of selectivity in the system where either $\Sigma_s \geq 1$ or more likely $\Sigma_s \gg 1$. In contrast, the GHK model assumes that the effects of the surface charge density are negligible and that there is virtually no selectivity at all, such that $\Sigma_s \ll 1$. In fact, GHK explicitly assumes that $\Sigma_s = 0$. Thus, RED and GHK are supposedly dichotomic to each other. This dichotomy will be bridged shortly by deriving a solution for arbitrary $\Sigma_s$.

## 2. Zero-current voltage, $\tilde{V}_{\tilde{i}=0}$

We return to the two intercept points and start with the simpler of the two: the zero-current voltage, $\tilde{V}_{\tilde{i}=0}$. It has long been known that the Nernst potential yields the predicted potential drop across an ideally selective system subjected to a concentration gradient

$$\tilde{V}_{\text{Nernst}} = \tilde{V}_{th} \ln\left(\frac{\tilde{c}_{\text{right}}}{\tilde{c}_{\text{left}}}\right), \tag{3}$$

where $\tilde{V}_{th}$ is thermal potential (and will be defined shortly).

The goal of GHK was to extend this equation for multiple species of arbitrary diffusion coefficients at vanishing selectivity ($\Sigma_s = 0$). In his work, Goldman derived two expressions for $\tilde{V}_{\tilde{i}=0}$. The first, which is highly ignored (Goldman's Eq. 13) is

$$\tilde{V}_{\text{Goldman}} = \tilde{V}_{th} \frac{\tilde{D}_+ - \tilde{D}_-}{\tilde{D}_+ + \tilde{D}_-} \ln\left(\frac{\tilde{c}_{\text{right}}}{\tilde{c}_{\text{left}}}\right), \tag{4}$$

where $\tilde{D}_+$ and $\tilde{D}_-$ are the diffusion cofficents of the positive and negative species. Three comments regarding notations are needed. First, throughout this work, we will compare our mathematical results to Goldman's [27]. For clarity, all of Goldman's equations will be numbered as Eq. (G#), where # will refer to the number of the particular equation from his work [27]. For example, Eq. (4) is Eq. (G13). Second, our notations differ from those used in Goldman's original work. These changes have been made to improve the clarity as well as to better reflect the standard notations used today (eighty years after his original work). Table 1 summarizes the differences in the notation and nomenclatures between our work and Goldman's. Finally, throughout this work, all quantities with tildes are dimensional, while quantities without tildes are dimensionless.

**Table 1**. Comparison of the notations used in this work versus that of Goldman [27]. For completion, we provide both our dimensional and dimensional-less variables.

| | Goldman [27] dimensional | This work dimensional | This work dimensionless |
|---|---|---|---|
| Concentrations and electric potential fields | $n_+, n_-, V$ | $\tilde{c}_+, \tilde{c}_-, \tilde{\phi}$ | $c_+, c_-, \phi$ |
| Space charge density | $\rho$ | $\tilde{\rho}_e$ | $\rho_e$ |
| Diffusion | $\alpha_\pm$ | $\tilde{D}_\pm$ | $D_- = \tilde{D}_+ / \tilde{D}_-$ |
| Mobility | $u_\pm = \dfrac{F\alpha_\pm}{RT}$ | Not used here | |
| Fixed charges or excess counterion concentration[1,2] | $\tilde{N}$ | $\tilde{\Sigma}_s$ | $\Sigma_s$ |
| Total concentration to mobile ions[2] | $N$ | $\tilde{C} = 2\tilde{c}$ | |
| Concentrations at the two ends of the channel | $n_\pm^0, n_\pm'$ | $\tilde{c}_{\text{left}}, \tilde{c}_{\text{right}}$ | $c_{\text{left}}, c_{\text{right}}$ |
| Potential drop | $-\Delta V$ | $\tilde{V}$ | $V$ |
| Thermal potential ($\tilde{V}_{th} = 1/\beta$) | $\beta = \tilde{F}/\tilde{R}\tilde{T}$ | $\tilde{V}_{th} = \tilde{R}\tilde{T}/\tilde{F}z$ | |
| Integration constants for governing equations[3] | $B, -gB$ | $-\tilde{\hat{j}}, -\tilde{\hat{i}}$ | $-\hat{j}, -\hat{i}$ |

[1] Our definition differs from that of Goldman's by a minus sign.

[2] Goldman uses $N$ and $\tilde{N}$ for two very different quantities. We use different notations for these quantities.

[3] In Goldman, the integration constants do not have physical meaning, while in this work, which focuses on two species, their physical meaning is clear. See text for further discussion.

The second and more famous expression Goldman derived, under the assumption of a uniform electric field, is given by [Eq. (G15)]

$$\tilde{V}_{\text{Goldman-uniform}} = \tilde{V}_{th} \ln\left(\frac{\tilde{D}_+ \tilde{c}_{\text{right}} + \tilde{D}_- \tilde{c}_{\text{left}}}{\tilde{D}_+ \tilde{c}_{\text{left}} + \tilde{D}_- \tilde{c}_{\text{right}}}\right). \qquad (5)$$

Note that when $\tilde{D}_+ = \tilde{D}_-$ or $\tilde{c}_{\text{right}} = \tilde{c}_{\text{left}}$ (or both), Eqs. (4) and (5) predict the same result (i.e., $\tilde{V}_{\text{Goldman-uniform}} = \tilde{V}_{\text{Goldman}} = 0$), else the results are different.

Equation (5) was extended both by Goldman [27] and Hodgkins and Katz [28] to hold for multiple species. This famous form is then given succinctly by [Eq. (G18)]

$$\tilde{V}_{\text{Goldman-uniform-multiple}} = \tilde{V}_{th} \ln\left(\frac{\Lambda_-}{\Lambda_+}\right), \tag{6}$$

where $\Lambda_\pm$ are functions of the concentrations at the two ends of the systems. Typically, one finds that instead of using $D_\pm$, $\Lambda_\pm$ uses the related permeabilities. Since we will soon prove that Eq. (5), and subsequentially Eq. (6), is inappropriate, then we will not provide the rather long and tedious expressions for $\Lambda_\pm$.

It is essential to discuss the differences between Eqs. (4) and (5). Since they provide two completely different models for $\tilde{V}_{\tilde{i}=0}$ (when $\Sigma_s = 0$), several important questions arise: 1) How different are Eqs. (4) and (5)? 2) How can there be two different equations describing the same thing? Can both be "correct"? And if not, which is the "correct" one? 3) Can the "correct" equation somehow be related to the Nernst equation [Eq. (3)] for very large values of $\Sigma_s \gg 1$?

The first question is the easiest to answer. Figure 4 plots the ratio of the two equations for LiCl, NaCl, and HCl for a given value of $\tilde{c}_{\text{left}} = 10^{-3} [\text{mol}/\text{m}^3]$ vs $\tilde{c}_{\text{right}}$ (all diffusion coefficients are given in Table 2). Depending on the system parameters, the ratio can be larger than a factor of 2 (i.e., 100% error). The answer to the second set of questions is slightly more difficult (and frustrating). It will be argued (in Secs. II.C.3 and III.A-B) that Eqs. (4) and (5) represent two different solutions that are non-commutative since they embed substantially different assumptions. Further, within their appropriate mathematical limits, they are both correct. However, as we will demonstrate shortly (using non-approximated numerical simulations), any quantity derived from the assumption of a uniform electric field (i.e., a linear potential drop) does not comply with the relevant physics. The final question will also be resolved. We will show that our new general solution holds for arbitrary $\Sigma_s$ and can reproduce Eq. (3) and Eq. (4) are the limits of $\Sigma_s \gg 1$ and $\Sigma_s = 0$, respectively.

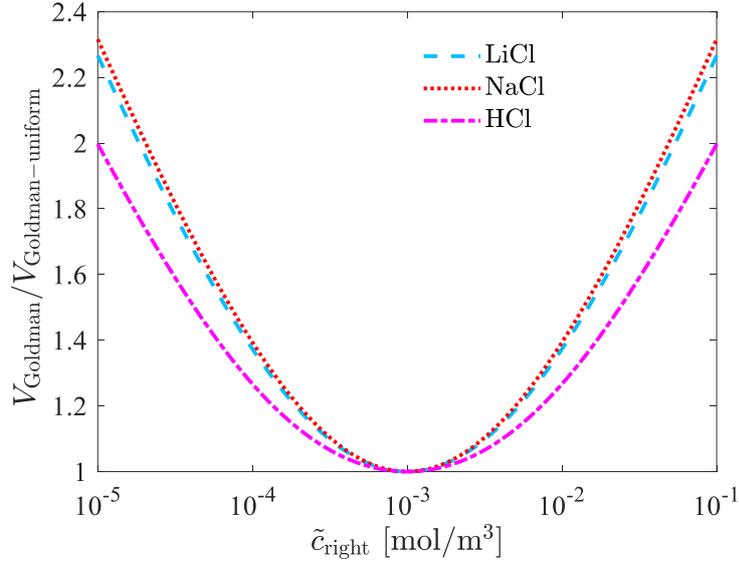

**Figure 4**. The ratio of $\tilde{V}_{\text{Goldman}} / \tilde{V}_{\text{Goldman-uniform}}$ vs. $\tilde{c}_{\text{right}} / \tilde{c}_{\text{left}}$ for LiCl, NaCl, and HCl for $\tilde{c}_{\text{left}} = 10^{-3} [\text{mol}/\text{m}^3]$. All diffusion coefficients are given in Table 2

## 3. Zero-voltage current, $\tilde{i}_{\tilde{V}=0}$

The most important characteristic of RED is the zero-voltage current, $\tilde{i}_{\tilde{V}=0}$. This current is the harvestable current in RED. It is related to the available power density, $\tilde{P}$, that can be harvested in RED. However, there are two common and related misconceptions about how to calculate $\tilde{i}_{\tilde{V}=0}$ and $\tilde{P}$.

The power density of metallic conductor is given by $\tilde{P} = \tilde{i}\tilde{V} = \tilde{V}^2 \tilde{g}_{\text{Ohmic}} = \tilde{i}^2 \tilde{g}_{\text{Ohmic}}^{-1}$. The transition from one expression to the other assumes that no matter the circumstance, the current-voltage response is always linear. While such a statement might hold for a metallic conductor, it does not hold for a nanofluidic system where the response is inherently nonlinear and is shifted from the origin. In particular, Figure 2 demonstrates that the asymmetric concentration curve is not a straight line (as is the curve of the symmetric concentration), where it can be observed that $\tilde{g}_{\text{Ohmic}} = \tilde{g}_{\tilde{i}=0} = (d\tilde{i}/d\tilde{V})_{\tilde{i}=0}$ is not equal to $\tilde{g}_{\tilde{V}=0} = (d\tilde{i}/d\tilde{V})_{\tilde{V}=0}$.

Moreover, the change in the local (differential) conductance and the shift from the origin present a conceptual problem related to how to calculate the power density. Based on the definition $\tilde{P} = \tilde{i}\tilde{V}$, the power density at the two intercepts ($\tilde{i} = 0$ or $\tilde{V} = 0$) is identically zero. However, this cannot be the case since RED harvests energy/power for zero applied voltage. The two other definitions, $\tilde{P}_s = \tilde{V}_s^2 \tilde{g}_s$ or $\tilde{P}_s = \tilde{i}_s^2 \tilde{g}_s^{-1}$ circumvent this problem. Here, we added the subscript '*s*' to ensure that it is clear that both the current and the voltage must be in the same state of either $\tilde{i}=0$ or $\tilde{V}=0$. These two states are inherently different (physically and mathematically). One definition yields the power that is needed to create a potential difference across the channel – this is $\tilde{P}_{\tilde{i}=0}=\tilde{V}_{\tilde{i}=0}^2\tilde{g}_{\tilde{i}=0}$. The other yields the power density that can be harvested from RED – this is $\tilde{P}_{\tilde{V}=0}=\tilde{i}_{\tilde{V}=0}^2\tilde{g}_{\tilde{V}=0}^{-1}$. These two expressions are different than those that are commonly found in the literature, $\tilde{P}=\tilde{i}_{\tilde{V}=0}\tilde{V}_{\tilde{i}=0}$ which often utilizes the relation $\tilde{i}_{\tilde{V}=0}=\tilde{V}_{\tilde{i}=0}\tilde{g}_{\text{Ohmic}}$ (here $\tilde{g}_{\text{Ohmic}}=\tilde{g}_{\tilde{i}=0}$). It is easy to observe that $\tilde{P}=\tilde{i}_{\tilde{V}=0}\tilde{V}_{\tilde{i}=0}$ utilizes properties from two different states. This is strictly incorrect and can lead to the incorrect estimation of the power density. In Sec. V.C, we will discuss this further and demonstrate how to calculate $\tilde{i}_{\tilde{V}=0}$ and $\tilde{P}_{\tilde{V}=0}=\tilde{i}_{\tilde{V}=0}^2\tilde{g}_{\tilde{V}=0}^{-1}$ in a self-consistent manner.

### C. Governing equations

#### 1. Dimensional form

In general, most electrolytes are comprised of several species. This is, of course, true of bodily fluids modeled by GHK, as well as seawater used in RED. In fact, the final objective of GHK was to provide a detailed characteristic of all transport phenomena for an electrolyte with an arbitrary number of ionic species. In this work, we will focus only on an electrolyte comprised of two species: a positive species and a negative species. This two-species electrolyte will be enough to demonstrate that GHK is inappropriate. Further, the mathematics that will be used shortly will be challenging enough that it is preferable that we focus on the much simpler problem.

The 1D steady-state convectionless transport of a two-species electrolyte through a charged nanochannel can be modeled with the (dimensional) Poisson-Nernst-Planck equations

$$-\partial_{\tilde{x}}\cdot\tilde{j}_{\pm}=\partial_{\tilde{x}}\left(\tilde{D}_{\pm}\partial_{\tilde{x}}\tilde{c}_{\pm}\pm\frac{z_{\pm}\tilde{D}_{\pm}\tilde{F}}{\tilde{R}\tilde{T}}\tilde{c}_{\pm}\partial_{\tilde{x}}\tilde{\phi}\right)=0, \tag{7}$$

$$\tilde{\varepsilon}_0\varepsilon_r\partial_{\tilde{x}\tilde{x}}\tilde{\phi}=-\tilde{\rho}_e=-\tilde{F}(z_+\tilde{c}_+-z_-\tilde{c}_--\tilde{\Sigma}_s). \tag{8}$$

Here, $\tilde{R}$ is the universal gas constant, $\tilde{T}$ is the absolute temperature (assumed to be uniform across the system), $\tilde{F}$ is the Faraday constant, $z_{\pm}=\pm z$ are the ionic valences, $\tilde{D}_{\pm}$ are the diffusion coefficients, and $\tilde{\varepsilon}_0$ and $\varepsilon_r$ are the permittivity of free space and the relative permittivity, respectively. The Nernst-Planck equations [Eq. (7)] govern ion flux conservation for a positive and negative species, $\tilde{c}_+$ and $\tilde{c}_-$, respectively. The flux densities for the positive and negative species are given by $\tilde{j}_+$ and $\tilde{j}_-$, respectively. The Poisson equation [Eq. (8)] governs the electrical potential, $\tilde{\phi}$, and depends on the space charge density, $\tilde{\rho}_e$. Finally, we note that $\tilde{\Sigma}_s$ is the excess counterion concentration due to the surface charge (or as Goldman termed them, these are the "fixed charges"). For a nanochannel system, $\tilde{\Sigma}_s = -\tilde{\sigma}_s \tilde{P}/(\tilde{F}\tilde{A})$ where $\tilde{\sigma}_s$ is the surface charge density, $\tilde{P}$ is the perimeter of the nanochannel's cross-section with area $\tilde{A}$.

We already note several small differences in our Eqs. (7)-(8) to those of Goldman's Eq. (G1)-(G3). Goldman's Nernst-Planck equation [Eq. (G1)] has a superfluous valency, $z_i$, and Faraday constant, $F$, multiplying the bracketed term in Eq. (7). Goldman's Poisson equation [Eq (G2)] has $4\pi$ that we do not have in our Eq. (8). This can be ignored for two reasons. First, he used a different system of units. Second, as we will shortly demonstrate, once the governing equations are normalized, the effect of the dielectric appears only in the Debye length. However, soon, similar to Goldman, we will assume that the problem is electroneutral such that dependency on the Debye length vanishes, and with it, the dependency on $4\pi$.

To complete the problem formulation, we specify the conditions at the edge of the system. At the two ends of the system, we have two different bulk concentrations and a total potential drop, $\tilde{V}$

$$\begin{aligned} \tilde{\phi}(\tilde{x}=0) = \tilde{V}, & \quad \tilde{c}_\pm(\tilde{x}=0) = \tilde{c}_{\text{left}} \\ \tilde{\phi}(\tilde{x}=\tilde{L}) = 0, & \quad \tilde{c}_\pm(\tilde{x}=\tilde{L}) = \tilde{c}_{\text{right}} \end{aligned}. \qquad (9)$$

Note that we will not term these boundary conditions just yet as there is a very important subtlety regarding the difference between edge conditions and the (correct) boundary conditions that should be applied (or, to be more exact, how the edge conditions need to be applied). These subtleties will be discussed in the next sections when we rederive GHK results and when we derive the new solution.

Importantly, it should become evident that Eqs. (7)-(9) are identical in both GHK and RED systems. In other words, and perhaps unsurprisingly, ion channels are miniature RED pumps. If we have learned anything from the past, they are also likely going to be the most efficient RED systems.

## 2. Governing equations – non-dimensional form

To streamline the derivation, it is beneficial to rewrite all equations in non-dimensional form. We shall use the following normalizations

$$x = \frac{\tilde{x}}{\tilde{L}}, \phi = \frac{\tilde{\phi}}{\tilde{V}_{th}}, c_{\pm} = \frac{\tilde{c}_{\pm}}{\tilde{c}_{bulk}}, \rho_e = \frac{\tilde{\rho}_e}{\tilde{F}\tilde{c}_{bulk}}, j_{\pm} = \frac{\tilde{j}_{\pm}}{\tilde{j}_0}. \tag{10}$$

All lengths and ensuing derivatives are normalized by the nanochannel length $\tilde{L}$, which in the reduced 1D model is the only remaining length scale. In the remainder, even though $L = 1$, we keep $L$ in all expressions to improve the clarity regarding the role of $L$ (or $\tilde{L}$). The electric potential and potential drop, $V$, are normalized by the thermal potential, $\tilde{V}_{th} = \tilde{R}\tilde{T}/\tilde{F}z$. Here $\tilde{c}_{bulk}$ is either of the bulk reservoirs (i.e., one of the bulk concentrations is set to unity). The space charge density has been normalized by $\tilde{F}\tilde{c}_{bulk}$. The ionic currents are normalized by $\tilde{j}_0 = \tilde{D}_+ \tilde{c}_{bulk}/\tilde{L}$. Here, we have arbitrarily chosen to normalize using the positive diffusion coefficient and normalize the negative diffusion coefficient such that $D_- = \tilde{D}_-/\tilde{D}_+$ which will appear in many of the expressions below. Also, note that the electrical current density [to be defined shortly, Eq. (20)] will be normalized by $\tilde{F}\tilde{j}_0$. Upon insertion of Eq. (10) into Eqs. (7)-(9), we have

$$-\partial_x j_+ = \partial_x (\partial_x c_+ + c_+ \partial_x \phi) = 0, \tag{11}$$

$$-\partial_x j_- = \partial_x [D_- (\partial_x c_- - c_- \partial_x \phi)] = 0, \tag{12}$$

$$2\lambda_D^2 \partial_{xx} \phi = -\rho_e = -(c_+ - c_- - \Sigma_s). \tag{13}$$

$$\begin{aligned} \phi(x=0) = V, \quad c_{\pm}(x=0) = c_{left} \\ \phi(x=L) = 0, \quad c_{\pm}(x=L) = c_{right} \end{aligned}. \tag{14}$$

Here, $\lambda_D = \tilde{\lambda}_D/\tilde{L} = \sqrt{\varepsilon_0 \tilde{\varepsilon}_r \tilde{V}_{th}/2\tilde{F}\tilde{c}_{bulk}}/\tilde{L}$ is the normalized Debye length [commonly termed the electric double layer (EDL)] and $\Sigma_s = \tilde{\Sigma}_s/\tilde{c}_{bulk}$ is a non-dimensional number/parameter that has two interpretations: the normalized average excess counterion concentration (Goldman's "fixed charges") or the Duhkin number [46](the ratio of the conductance associated with the surface charge and the conductance associated with the bulk). It will become evident that the limit $\Sigma_s = \tilde{\Sigma}_s/\tilde{c}_{bulk} \gg 1$ corresponds to the case that the bulk concentrations are very small compared to the effects of the surface charge. In the opposite case, $\Sigma_s \ll 1$ corresponds to the case that bulk concentration effects are predominant (vanishing selectivity). A discussion regarding $\Sigma_s$ and $\lambda_D$ and their interrelation is provided below.

### 3. Reduced governing equations

Equations (11)-(13) can be further simplified. Requiring conservation of fluxes, Eqs. (11)-(12) can be integrated to yield

$$-j_+ = \partial_x c_+ + c_+ \partial_x \phi, \qquad (15)$$

$$-j_- = D_-(\partial_x c_- - c_- \partial_x \phi), \qquad (16)$$

where $j_-$ and $j_-$ remain unknown constants of integration that remain to be determined.

There are two approaches to simplifying Eq. (13). Both approaches are utilized by Goldman. First, it can be assumed that the EDL is sufficiently small such that $\lambda_D^2 \partial_{xx} \phi$ can be taken to zero. This yields the constraint of electroneutrality whereby $\rho_e = 0$ or $c_+ - c_- - \Sigma_s = 0$. Note that also Goldman utilizes $\rho_e = 0$ [his statement is given above Eq. (G4)]. The second approach used by Goldman (but not adopted here) is the famous GHK assumption of a uniform electric field $E = V/L$, which yields

$$\phi_{GHK} = V(1-x)/L. \qquad (17)$$

This potential satisfies $\partial_{xx} \phi = 0$ for arbitrary values of $\lambda_D$.

Both approaches are mathematically valid. In the former approach, one requires that the right-hand of Eq. (13) is zero – implying that $\lambda_D^2 \partial_{xx} \phi$ must be a small number. In the latter approach, one requires that the left-hand side of Eq. (13) is zero, while the other implies that $\rho_e$ is a small but non-zero number. In other words, these two approaches differ in that they are different approximations: electroneutrality or a linear potential drop. Notably, the approximation depends on whether $\lambda_D \ll 1$ (the electroneutrality approach) or $\lambda_D \gg 1$ (the GHK approach). However, the limit $\lambda_D \gg 1$ is precarious – for $\Sigma_s$ depends on $\lambda_D$ in such a manner that it is difficult to assume that $\lambda_D \gg 1$ unless $\Sigma_s \triangleq 0$ (however, then one is over-specifying the problem as both electroneutrality and uniform electric field are assumed).

Consider the case of a cylinder of radius $\tilde{a}$ ($\tilde{P} = 2\pi\tilde{a}$ and $\tilde{A} = \pi\tilde{a}^2$) with $\tilde{\Sigma}_s = -2(\sigma_s \tilde{\sigma}_d)/(\tilde{F}\tilde{a})$ where the non-dimensional surface charge density $\sigma_s = \tilde{\sigma}_s / \tilde{\sigma}_d$ is normalized [47] by $\tilde{\sigma}_d = \tilde{\varepsilon}_0 \varepsilon_r \tilde{V}_{th} / \tilde{a}$. Then, $\Sigma_s = \tilde{\Sigma}_s / \tilde{c}_{\text{bulk}} = -4\sigma_s(\tilde{\lambda}_D^2/\tilde{a}^2) = -4\sigma_s \lambda_D^2 (\tilde{L}^2/\tilde{a}^2)$ is a function of $\lambda_D$. In order for $\Sigma_s \triangleq 0$, one must require that either $\sigma_s \triangleq 0$ or that $\lambda_D \triangleq 0$ (or both). We shall not consider the case $\tilde{\lambda}_D \triangleq 0$ for three reasons: 1) $\lambda_D$ is never truly zero; 2) $\lambda_D \ll 1$ is opposite to the GHK approach; 3) Eq. (13) would become singular. Thus, we consider the very unphysical situation that $\sigma_s \triangleq 0$.

In the GHK scenario when $\Sigma_s \triangleq 0$ (and not $\Sigma_s \approx 0$ as described above), there is no natural symmetry breaking between positive and negative ions anywhere in the system leading to $c_+ = c_- = c$. Inserting $c_+ = c_- = c$ and Eq. (17) into Eqs. (15)-(16), and taking their difference [after multiplying Eq. (15) by $D_-$] yields $\partial_x \phi = -(D_- j_+ - j_-)/(2D_- c)$. This expression can be a constant only if $c$ is a constant. However, it is trivial to show that this is not the case. Taking the sum of Eqs. (15)-(16) [after multiplying Eq. (15) by $D_-$], yields $\partial_x c = -(j_+ + j_-/D_-)$. Unless, $\partial_x c = 0$ is zero (and it is generally not), such that $c$ is a constant the $\partial_x \phi \neq const$. In other words, we already find that Eq. (17) can satisfy the requirement that $\lambda_D^2 \partial_{xx} \phi = 0$, but the final result will not yield a self-consistent result on how the behavior of the electric field.

Since, in reality, the surface charge is never truly zero, and, by definition, the EDL is never zero (both can be small, but they are finite), we must reconsider the assumption that $\Sigma_s \triangleq 0$. Instead, we will adopt the approach that utilizes the $c_+ = c_- + \Sigma_s$ – it is here that our approach will differ from GHK.

Importantly, regardless of the approach, these simplified equations [Eqs. (15)-(16)] are still subject to the edge conditions given in Eq. (14). When $\Sigma_s \neq 0$, we will modify Eq. (14) to account for the Donnan potential drop.

## III. THE GHK MODEL
### A. Goldman's derivation

In the following, we shall further demonstrate that Goldman's derivation, while mathematically allowed, is internally inconsistent. To that end, we will focus on two species (and not an arbitrary number of species) that are symmetric whereby $\tilde{D}_- = \tilde{D}_+$ (such that $D_- = 1$.) These two simplifying assumptions are allowed as Goldman's approach is not singular for two salts or even for a KCl-like electrolyte.

Following Goldman, we define the "total concentration to mobile ions"

$$C = c_+ + c_-. \qquad (18)$$

Goldman then adds and subtracts the two flux equations [Eq. (15)-(16)], while accounting for $c_+ - c_- - N = 0$, yielding [Eqs (G4)-(G5)]

$$-\hat{j} = \partial_x(C + \Sigma_s) + \Sigma_s \partial_x \phi = \partial_x C + \Sigma_s \partial_x \phi, \tag{19}$$

$$-\hat{i} = \partial_x \Sigma_s + (C + \Sigma_s)\partial_x \phi = (C + \Sigma_s)\partial_x \phi, \tag{20}$$

where

$$\hat{j} = j_+ + j_-, \quad \hat{i} = j_+ - j_-, \tag{21}$$

are the salt current and electrical current densities given, respectively. Several comments regarding Eq. (19)-(21) are essential. First, we have leveraged that $\Sigma_s$ is spatially independent (i.e., $\partial_x \Sigma_s = 0$). Second, observe that $\Sigma_s$ appears in both Eqs.(19) and (20). In contrast to this work, Goldman keeps $\Sigma_s$ in his Eq. (G5) [our Eq. (19)] but removes it from his Eq. (G4) [our Eq. (20)]. While this is not strictly rigorous, perhaps it is permissible since his analysis focuses entirely on the $\Sigma_s = 0$ scenario. In the remainder of the rederivation of his results, when possible, we will keep this term. When it becomes overwhelmingly burdensome to keep, it shall be ignored. Third, at this point, like $j_+$ and $j_-$, both $\hat{i}$ and $\hat{j}$ are still unknown constants of integration. Fourth, in Goldman's notation, these two constants of integration ($\hat{i}$ and $\hat{j}$) are not recognized as the flux densities. This is because Goldman considers an electrolyte comprised of several salts with different diffusion coefficients (or mobilities), such that in his work, the equivalent of $\hat{i}$ and $\hat{j}$ are not truly the salt current and electrical current densities. Hence, we have marked them with a hat to differentiate them from the salt current and electrical current densities $i$ and $j$, which will be defined in the next section. In the next section that deals with $\tilde{D}_- \neq \tilde{D}_+$, we will define flux densities $i$ and $j$ differently than $\hat{i}$ and $\hat{j}$. However, importantly, when $\tilde{D}_- = 1$, $i = \hat{i}$ and $j = \hat{j}$.

Goldman integrates Eq. (19) using two different approaches. The indefinite integral yields the total concentration drop as a function of $x$ [Eq. (G6)]

$$C(x) = C_{\text{left}} - \hat{j}x - \Sigma_s[\phi(x) - V]. \tag{22}$$

Here, we have used only one concentration edge condition from using Eq. (14). While the definite integral between the two ends of the channel [using Eq. (14)] yields the salt current flux density [Goldman's Eq. (G7)]

$$-\hat{j}L = (C + \Sigma_s \phi)\big|_{x=0}^{x=L} \Rightarrow \hat{j} = \frac{C_{\text{left}} - C_{\text{right}} + \Sigma_s V}{L}, \tag{23}$$

Shortly, we will use two relations[Eqs. (22)-(23)] when $\Sigma_s = 0$. First, in the derivation of Eq. (26) below, we will leverage Eq. (23) and rewrite it as $C_{\text{right}} = C_{\text{left}} - \hat{j}L$. Second, in the comparison at the end of the section, we will also leverage that for salts with the same valency [$z_\pm = \pm z$, see Eq. (33)], Eq. (22) can be rewritten for each concentration

$$c_\pm = c = c_{\text{left}} - \tfrac{1}{2}\hat{j}x. \tag{24}$$

Goldman's approach to calculating the electric potential seems rather elaborate. For the sake of simplicity, we will take a simpler and more straightforward approach that recapitulates his result. We will insert $\Sigma_s = 0$ into Eq. (22) and insert the resultant expression into Eq. (20), allowing for a direct and simple integration of the potential

$$-\hat{i} = (C_{\text{left}} - \hat{j}x)\partial_x \phi \Rightarrow \phi = \frac{\hat{i}}{\hat{j}}\ln(C_{\text{left}} - \hat{j}x) + const. \tag{25}$$

We can easily find the constant by requiring $\phi(x=L) = 0$ such that [this is Eq. (G8) upon inserting $\Sigma_s = 0$ into Eq. (G8)]

$$\phi = \frac{\hat{i}}{\hat{j}}\ln\left(\frac{C}{C_{\text{right}}}\right) = \frac{\hat{i}}{\hat{j}}\ln\left(\frac{C_{\text{left}} - \hat{j}x}{C_{\text{right}}}\right). \tag{26}$$

The constraint, $\phi(x=0) = V$, leads to a current-voltage, $i - V$, response

$$V = \frac{\hat{i}}{\hat{j}}\ln C_r \Rightarrow \hat{i} = \frac{V\hat{j}}{\ln(C_{\text{left}}/C_{\text{right}})} = \frac{C_{\text{left}} - C_{\text{right}}}{\ln(C_{\text{left}}/C_{\text{right}})}\frac{V}{L}. \tag{27}$$

We note that this equation is what would be calculated from Goldman's Eq. (G12) for the case of a symmetric diffusion coefficient. However, Eq. (G12) is all but ignored in all future derivations for electrolytes of multiple species and is replaced by a constant electric field approximation – discussed in the subsection below.

### B. Uniform electric field approximation

In general, the $i-V$ is the end goal of any derivation of ion transport as it fully characterizes a system after accounting for all systems parameters. Ideally, Goldman should have stopped his derivation here, but instead, he embarks on a different path where he inserts his famous constant electrical field assumption. Goldman [27] states below his Eq. (G10) [in our notation]

> "When $\tilde{\Sigma}_s = 0$ (the Planck case), the potential distribution is logarithmic and the total concentration is linear although the individual ion concentrations are not. If

> $\tilde{\Sigma}_s$ *is not zero, the potential distribution is more complicated and the total concentration distribution is no longer linear. Evidently the current-voltage relation is, in general, nonlinear as well."*

The results we have just derived agree with this statement. However, without proper justification, Goldman then hypothesizes that a constant electrical field can be used

> *"We then approach a situation in which the field is constant and are led to a solution analogous to that given by Mott (1939) [48] for electronic conduction in the copper-copper oxide rectifier."*

It is worthwhile to quote Hodgin and Katz [28], whose assumptions are similar to Goldman's. In their Appendix, they write:

> *The basic assumptions are (1) that ions in the membrane move under the influence of diffusion and the electric field in a manner which is essentially similar to that in free solution; (2) that the electric field may be regarded as constant throughout the membrane; (3) that the concentrations of ions at the edges of the membrane are directly proportional to those in the aqueous solutions bounding the membrane; and (4) that the membrane is homogeneous.*

Assumption (1) is a statement that $\Sigma_s = 0$. Assumption (2) repeats the uniform electric field assumption. Assumption (3) is an assumption on the edge condition [Eq. (9)]. Finally, assumption (4) is that fluxes are uniform over the cross-section (equivalent to assuming 1D). These assumptions are equivalent to Goldman's, making both models equivalent.

We insert Eq. (17) [i.e. $E = V/L$] into Eqs. (15)-(16)

$$-j_+ = \partial_x c_+ - c_+ V/L, \tag{28}$$

$$-j_- = \partial_x c_- + c_- V/L. \tag{29}$$

It is trivial to show that when one uses the conditions of $c_\pm(x=0) = c_{\text{left}}$, integration of Eqs. (28)-(29) yields the two *exponential* concentration distributions

$$c_{\text{GHK},+} = \frac{j_+ L}{V} + \left(c_{\text{left}} - \frac{j_+ L}{V}\right)e^{\frac{Vx}{L}}, \quad c_{\text{GHK},-} = -\frac{j_- L}{V} + \left(c_{\text{left}} + \frac{j_- L}{V}\right)e^{\frac{-Vx}{L}}. \tag{30}$$

Requiring that $c_\pm(x=L) = c_{\text{right}}$ yields the fluxes [Eq. (G11)]

$$j_{\text{GHK},+} = \frac{c_{\text{right}}e^{-V} - c_{\text{left}}}{e^{-V} - 1}\frac{V}{L}, \quad j_{\text{GHK},-} = \frac{c_{\text{left}} - c_{\text{right}}e^{V}}{e^{V} - 1}\frac{V}{L}, \tag{31}$$

which yields the salt current density and electric current density [this is tantamount to assuming equal mobilities in Eq. (G14)],

$$\hat{j}_{\text{GHK}} = j_{\text{GHK},+} + j_{\text{GHK},-} = \Delta C \frac{V}{L}\coth\left(\frac{V}{2}\right), \quad \hat{i}_{\text{GHK}} = j_{\text{GHK},+} - j_{\text{GHK},-} = (C_{\text{right}} + C_{\text{left}})\frac{V}{L}. \tag{32}$$

Even before comparing to numerical simulations, there are several peculiarities with Eqs. (28)-(32). First, it is easy to observe that the total concentration, $C_{\text{GHK}} = c_{\text{GHK},+} + c_{\text{GHK},-}$, is not linear as rationalized by Goldman and calculated in Eq. (22). Second, the space-charge density $\rho_{e,\text{GHK}} = c_{\text{GHK},+} - c_{\text{GHK},-} \neq 0$ implying that $\Sigma_s \neq 0$. Also, global electroneutrality, $\int_0^L \rho_e dx$, is not zero (except for $V = 0$), whereas Goldman suggested that at the very least $\int_0^L \rho_e dx = 0$ (if not $\rho_{e,\text{GHK}} = 0$). Third, since the space charge density is not zero, the uniform electric field is not a self-consistent solution. Fourth, the salt current density, $\hat{j}_{\text{GHK}}$, is always positive and varies drastically with $V$. This is in contrast to Eqs. (23), which is linear with the voltage and can switch signs for sufficiently large enough voltages. Finally, the electrical current density $\hat{i}_{\text{GHK}}/V = (C_{\text{right}} + C_{\text{left}})/L$ yields a simple expression for the conductance. This expression is what would be expected if you add the contribution of two electrolytes at two different concentrations, but it doesn't appear to account for how the conductivity varies as a function of space [in contrast to Eq. (27)]. Perhaps this expression can be considered to be the upper limit to the conductance.

### C. Comparison of GHK to non-approximated numerical simulations

Figure 5 compares the key results predicted by the GHK model, derived in the previous subsection, to non-approximated numerical simulations. The details of the numerical procedure are given below in Appendix A).

Figure 5(a) presents the positive and negative concentration distribution ratios of numerical simulation to the exact profile [Eq. (24)] and the ratios of the GHK theory [for a uniform electric field given by Eq. (30)] to the exact profiles. The former has remarkable correspondence, while the latter has an approximate 5% difference (this difference depends on the voltage, concentration gradient, and more). Perhaps it can be argued that such a small difference is inconsequential and allowable. However, it should be pointed out that the fluxes $j_{\text{GHK},+}$ and $j_{\text{GHK},-}$ have been constrained to satisfy the concentration boundary conditions, which also constrains the behavior of the concentrations. Furthermore, we reiterate that if the concentrations are not identical, the space charge density is non-zero.

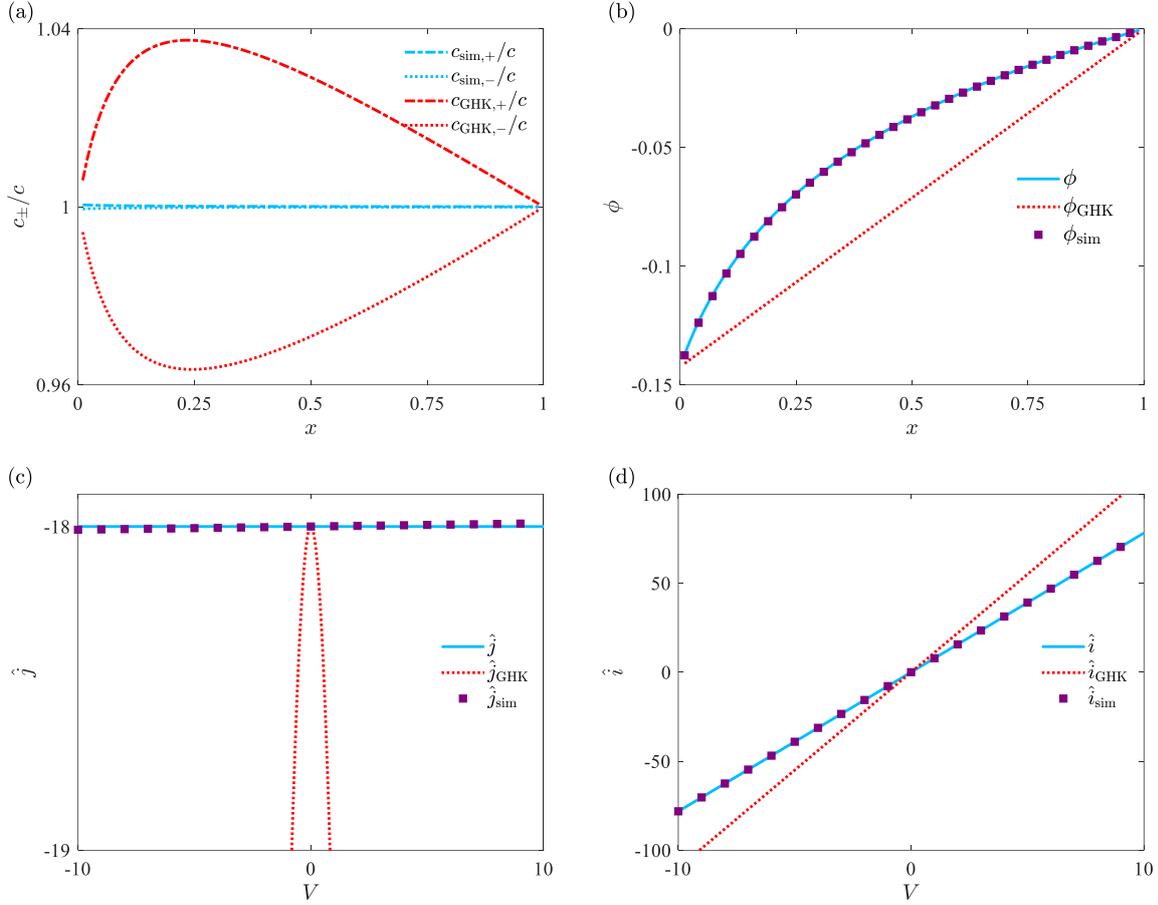

**Figure 5**. Comparison of the GHK model to an internally consistent model and exact numerical simulations for KCl ($D_- = \tilde{D}_K / \tilde{D}_{Cl} = 1$). (a) The ratio of the concentration calculated by GHK, $c_{GHK,\pm}$ [Eqs. (30)], or numerical simulation, to the exact concentration, $c$ [Eq. (24)]. (b) The potential distributions from numerical simulations, the linear drop predicted by GHK [Eq. (17)] and the exact solution [Eq. (26)]. (c-d) A comparison between the GHK model [Eq. (32)] for the (c) salt-current and (d) electric current density, numerical simulations, and the exact solutions [Eq. (23) and (27)]. The simulation parameters used in this figure are: $c_{right} = 10 c_{left}$ with $\tilde{c}_{left} = 1000 [\text{mol/m}^3] \gg \tilde{\Sigma}_s = 1[\text{mol/m}^3]$. Plots (a) and (b) are plotted for $V = -0.142$.

Figure 5(b) shows that the linear potential drop approximation [Eq. (17)] does not match simulations. In contrast, the logarithmic profile [Eq. (26)] fits perfectly. Similarly, and perhaps no longer unexpectedly, Figure 5 (c)-(d) show how $\hat{j}_{GHK}$ and $\hat{i}_{GHK}$ [Eq. (32)] do not match numerical simulations in contrast to the exact fluxes given by Eqs. (23) and (27). In particular, we see that $\hat{j}_{GHK}$ behaves drastically different than simulations. The theoretical solution for $\hat{j}$ does not match "perfectly", but then again, the theory was derived in the limit of $\Sigma_s = 0$, while numerical simulations were conducted for $\Sigma_s = 10^{-3}$. We also see in Figure 5 (d) that the Ohmic conductance is over-predicted by $\hat{i}_{GHK}$, while the exact theoretical model holds nicely. Here, the contribution of $\Sigma_s = 10^{-3}$ to the conductance is negligible.

### D. Preliminary summary and discussion of GHK

In this section, we have rederived the GHK model following Goldman's approach. For simplicity, we focused on the scenario $D_- = 1$. We compared Goldman's uniform electric field model to numerical simulations. We find that the uniform electric field model is internally inconsistent. Since there is nothing singular in the derivation with $D_- = 1$, then a natural outcome is that GHK is also inconsistent for $D_- \neq 1$. Similarly, since there is nothing singular in the derivation for two species, then, in a similar manner, the results of the uniform electric field GHK models for several species [Eq. (6)] is also questionable.

It is worthwhile to dwell on the origin of the inconsistency – the assumption of a constant electric field. The fact that the uniform electric field approximation is inappropriate should not come as a surprise. In fact, prior to his lengthy derivation, Goldman stated

> *We now have a set of equations and conditions from which, ideally, the current-voltage relation of the membrane may be derived. This set is not integrable in closed form as it stands however, and we are forced to introduce still further simplification if usable results are to be reached. The nature of the equations does suggest that, in general, a linear relation will not occur and that rectification will be more evident the greater the asymmetry of mobilities and concentrations. These same factors also determine the membrane potential and so the finding that the initial slope of the conductance-current curve increases with the membrane potential may be expected in any case.*

In other words, Goldman explains that without his additional ad-hoc assumptions, he is unable to proceed with the derivation of his general solution for an arbitrary number of species. Thus, he sought inspiration elsewhere – namely Mott's work [48]. The desire for a solution {in this case, a nonlinear $i-V$, [Eq. (G14)]}, which utilizes simple and elegant physical reasoning, is rather understandable but carries with it many dangers.

## IV.   NEW MODEL FOR ARBITRARY VALUES OF $\Sigma_s$ AND $D_-$

### A.  Salt current and electrical current densities

Our starting point will once again be Eqs. (14)-(16)

$$-j_+ = \partial_x c_+ + c_+ \partial_x \phi ,$$

$$-j_- = D_-(\partial_x c_- - c_- \partial_x \phi) ,$$

$$\phi(x=0) = V, \quad c_\pm(x=0) = c_{\text{left}}, \quad \phi(x=L) = 0, \quad c_\pm(x=L) = c_{\text{right}} ,$$

with the assumption of electroneutrality

$$c_+ - c_- - \Sigma_s = 0 ,$$

for arbitrary values of $\Sigma_s$ and $D_-$. Our end goal is to derive the $i-V$ relation. The price we shall pay for deriving the $i-V$ is in contrast to Goldman we will not be able to derive local expressions for the concentration and the electric potential distributions.

Instead of using the total local mobile concentration $C$ (used in Sec. III), we will use the notation of the single species concentration, $c$, which is more convenient [$C$ and $c$ are related through Eq. (18)]. Also, to further simplify the equations, we use the following notation (which removes the cumbersome $\pm$ subscripts)

$$c_- = c, \quad c_+ = c_- + \Sigma_s = c + \Sigma_s . \tag{33}$$

To derive the $i-V$ relation, we will need to compute the values of the salt and electrical current densities

$$j(D_-) = j_+ + j_-(D_-), \quad i(D_-) = j_+ - j_-(D_-) . \tag{34}$$

Even upon insertion of Eq. (33) into Eq. (34), the resultant equations, which depend on $D_-$, are rather complicated. The involved mathematical complexities are substantially more difficult than those involved with solving the equations for symmetric diffusion coefficients ($D_- = 1$).

These new difficulties can be circumvented if we divide each flux by its diffusion coefficient, and take the sum and difference,

$$\hat{j} = j_+ + j_-/D_-, \quad \hat{i} = j_+ - j_-/D_- . \tag{35}$$

We recover the equations that were previously solved in Sec. III (for $\Sigma_s = 0$) and in recently published work [33] (for arbitrary $\Sigma_s$). Note that these two equations [Eq. (35)], which for $D_- = 1$ are the real physical fluxes, are, in general, not the real physical fluxes. We shall term these "reduced" fluxes. Importantly, the real fluxes [Eq. (34)] and "reduced" fluxes [Eq. (35)] are related through their dependencies on $j_+$ and $j_-$ such that

$$\hat{j} = \frac{D_- - 1}{2D_-} i + \frac{1 + D_-}{2D_-} j \quad , \quad \hat{i} = \frac{1 + D_-}{2D_-} i + \frac{D_- - 1}{2D_-} j. \tag{36}$$

Knowledge of the "reduced" fluxes provides knowledge of the real fluxes (and vice versa). Our approach will be to reduce the real fluxes [Eq. (34)] to "reduced" fluxes [Eq. (35)] and solve the "reduced" governing equations for $\hat{j}, \hat{i}$ using our previous method [33]. Thereafter, we will transform to the real fluxes $j, i$ using Eq. (36).

Then, addition and subtractions of the fluxes in Eqs. (15)-(16) via Eq. (35), unsurprisingly, (re-)yields Eqs. (19)-(20)

$$-\hat{j} = \partial_x 2c + \Sigma_s \partial_x \phi, \tag{37}$$

$$-\hat{i} = (2c + \Sigma_s) \partial_x \phi. \tag{38}$$

Equation (38), which is a form of Ohm's law for the "reduced" fluxes (and is Ohm's law when $D_- = 1$), yields $\partial_x \phi = -\hat{i}/(2c + \Sigma_s)$. Substituting $\partial_x \phi$ into Eq. (37), and after rearranging, yields

$$\frac{dc}{dx} = \frac{1}{2} \left[ \frac{\hat{i} \Sigma_s - \hat{j}(2c + \Sigma_s)}{2c + \Sigma_s} \right], \tag{39}$$

which is a separable differential equation of the form [33]

$$\frac{2(2c + \Sigma_s)}{\hat{i} \Sigma_s - \hat{j}(2c + \Sigma_s)} dc = dx. \tag{40}$$

Eq. (40) can be integrated to yield [as long as $\hat{i} \neq \hat{j}(2c/\Sigma_s + 1)$]

$$\left\{ c + \frac{\hat{i} \Sigma_s}{2\hat{j}} \ln[(\hat{i} - \hat{j})\Sigma_s - 2c\hat{j}] \right\} \Bigg|_{c_{\text{nano-left}}}^{c_{\text{nano-right}}} = -\frac{\hat{j}L}{2}. \tag{41}$$

Note that $c_{\text{nano-left}} = c(x = 0^+)$ and $c_{\text{nano-right}} = c(x = L^-)$ are the unknown concentrations at the two ends of the nanochannel and are not to be confused with the two known conditions $c(x = 0^-) = c_{\text{left}}$ and $c(x = L^+) = c_{\text{right}}$. The latter two concentrations ($c_{\text{left}}$ and $c_{\text{right}}$) are the concentrations in the bulk reservoirs and can be thought of as the conditions that a distance of one EDL away from the boundaries into the bulk. In contrast, the two former concentrations ($c_{\text{nano-left}}$ and $c_{\text{nano-right}}$) can be thought of as conditions that are a distance of one EDL into the nanochannel. In the limit that the EDL length goes to zero, it is common (as Goldman did) to require that the concentrations and electric potential be continuous at the boundaries. This approach holds only when $\Sigma_s = 0$. However, if $\Sigma_s \neq 0$, a different approach is required.

Namely, as we transition from the reservoirs into the nanochannel, we do not require that the potential and concentrations be continuous. Instead, we require that the (non-dimensional) electrochemical potential

$$\mu_{\pm} = \ln c \pm \phi, \tag{42}$$

at the interface of the reservoirs and the nanochannel is continuous. This general and widely accepted approach will allow for jumps in the concentration and electric potential, which yield the established Donnan potential drop. Note that Eq. (42) represents four boundary conditions (two at each interface for two species). In the following, we will leverage two of the boundary conditions, while the remaining two will be used later when we derive a current-voltage response.

The sum of the positive and negative electrochemical potentials yields a simplified boundary condition that is independent of the potential

$$\mu = \mu_+ + \mu_- = \ln(c_+ c_-) \Rightarrow (c_+ c_-) = const. \tag{43}$$

Inserting Eq. (33) and Eq. (14) into Eq. (42) at $x = 0, L,$ yields $c_{\text{nano-}k}(c_{\text{nano-}k} + \Sigma_s) = c_k^2$ for $k = \text{left,right}$. The solution of the quadratic equation is

$$c_{\text{nano-}k=\text{left,right}} = -\tfrac{1}{2}\Sigma_s + S_k \quad , \quad S_k = \tfrac{1}{2}\sqrt{\Sigma_s^2 + 4c_k^2}. \tag{44}$$

We now insert $c_{\text{nano-}k=\text{left,right}}$ [Eq. (44)] into Eq. (41). This yields a relation between the reduced salt and reduced electrical current densities

$$f(\hat{i}, \hat{j}) = \frac{\hat{i}}{\hat{j}}\Sigma_s \ln\left(\frac{\hat{i}\Sigma_s - 2\hat{j}S_{\text{right}}}{\hat{i}\Sigma_s - 2\hat{j}S_{\text{left}}}\right) + \hat{j}L + 2(S_{\text{right}} - S_{\text{left}}) = 0. \tag{45}$$

To simplify the upcoming notation, we note that when the "reduced" electric current density is zero, $\hat{i} = 0$, Eq. (45), yields an intermediate "reduced" salt current density for zero reduced electric current density $\hat{j}_{i=0}$ [not to be confused with $j_{i=0}$ which will be derived and defined later – Eq. (56)]

$$\hat{j}_{i=0} = 2(S_{\text{left}} - S_{\text{right}})/L, \tag{46}$$

such that Eq. (45) can be written as

$$f(\hat{i}, \hat{j}) = \frac{\hat{i}}{\hat{j}} \Sigma_s \ln\left(\frac{\hat{i}\Sigma_s - 2\hat{j}S_{\text{right}}}{\hat{i}\Sigma_s - 2\hat{j}S_{\text{left}}}\right) + (\hat{j} - \hat{j}_{i=0})L = 0. \tag{47}$$

A detailed analysis of Eq. (47) will be provided shortly.

## B. Current-voltage response

The potential drop across the system is comprised of three terms

$$-V = \Delta\phi_{\text{Donnan-left}} + \Delta\phi_{\text{nano}} + \Delta\phi_{\text{Donnan-right}}. \tag{48}$$

Here, we have assumed that there are two additional Donnan potential drops across the system. Goldman correctly assumes that for $\Sigma_s = 0$, there are no Donnan potential drops. However, this is generally not the case. Here, we address the general case. We shall now calculate each of these term. We start with the simpler middle term and then calculate the two Donnan potential drops.

The middle term is the Ohmic drop across the nanochannel and can be found by integrating Eq. (37)

$$\Delta\phi_{\text{nano}} = -\frac{(\hat{j} - \hat{j}_{i=0})L}{\Sigma_s}. \tag{49}$$

The Donnan potential drops are calculated using the positive electrochemical potential at each interface ($x = 0, L$) yielding $(\ln c + \phi)_{x=0,L} = (\ln c_{\text{nano}-k} + \phi_{\text{nano}-k})_{k=\text{left,right}}$. Here, $\phi_{\text{nano}-k}$ are the potentials at the two ends of the nanochannel [similar to the previously used $c_{\text{nano}-k}$ terms in Eq. (41)]. The Donnan potential drops are

$$\Delta\phi_{\text{Donnan-left}} = \phi_{\text{nano-left}}(x=0^+) - \phi(x=0^-) = \ln\left[\frac{c(x=0)}{c_{\text{nano-left}} + \Sigma_s}\right] = \ln\left(\frac{c_{\text{left}}}{c_{\text{nano-left}} + \Sigma_s}\right), \tag{50}$$

$$\Delta\phi_{\text{Donnan-right}} = \phi(x=L^+) - \phi_{\text{nano-right}}(x=L^-) = \ln\left[\frac{c_{\text{nano-right}} + \Sigma_s}{c(x=L)}\right] = \ln\left(\frac{c_{\text{nano-right}} + \Sigma_s}{c_{\text{right}}}\right). \tag{51}$$

Inserting Eqs. (49)-(51) into Eq. (48), yields the total potential drop

$$V = \frac{(\hat{j} - \hat{j}_{i=0})L}{\Sigma_s} + \ln\left(\frac{c_{\text{right}}}{c_{\text{left}}}\right) - \ln\left(\frac{S_{\text{right}} + \frac{1}{2}\Sigma_s}{S_{\text{left}} + \frac{1}{2}\Sigma_s}\right). \tag{52}$$

It is important to note that Eq. (52) is still not the desired $i-V$. Rather, it is an intermediate $\hat{j}-V$ relation. Several additional steps will be needed to find the $i-V$. These are detailed in the following paragraphs. Before that, we would like to simplify the notation of Eq. (52) by noting that when $\hat{i} = 0$, the first term in Eq. (52) is also zero [due to Eq.(46)]. Then

$$\hat{V}_{\hat{i}=0} = \ln\left(\frac{c_{\text{right}}}{c_{\text{left}}}\right) - \ln\left(\frac{S_{\text{right}} + \frac{1}{2}\Sigma_s}{S_{\text{left}} + \frac{1}{2}\Sigma_s}\right), \tag{53}$$

such that Eq. (52) now has a very succinct and elegant form

$$V = (\hat{j} - \hat{j}_{i=0})\frac{L}{N} + \hat{V}_{\hat{i}=0}. \tag{54}$$

Similar to how Eq. (52) is still an intermediate relation, so too $\hat{V}_{\hat{i}=0}$ [Eq. (53)] is not the desired $V_{i=0}$ (which is the *measurable* voltage at zero current intercept with the *measured* $i-V$). Hence, one should not confuse $\hat{V}_{\hat{i}=0}$ with $V_{i=0}$ [which will be derived shortly in Eq. (64)].

### C. Transport number

To find an analytical expression for the $i-V$, several steps are needed, which we will briefly review before implementing them. To transform Eq. (54) from $\hat{j}-V$ to $i-V$, two steps are needed. First, we will insert Eq. (36), which transforms $(\hat{i}, \hat{j})$ to $(i, j)$, into Eq. (47), which transforms $f(\hat{i}, \hat{j}) = 0$ into $f(i, j) = 0$. Second, we will find a more convenient $i-j$ relation – we will transform $f(i, j) = 0$, again, to depend on the transport number, $\tau$, [or to be more exact, on the "differential transport number", $\Delta \tau$, defined below] such that $f(i, \Delta \tau) = 0$. Then, we take the Taylor series of $f(i, \Delta \tau) = 0$ at small currents to find the desired analytical expressions.

The transport number is defined as the contribution of the counterion flux to the total electrical current. For systems with symmetric diffusion coefficients and symmetric concentrations, this can be written as [49,50]

$$\tau = \frac{1}{2} + \frac{1}{2}\frac{j}{i} = \frac{1}{2} + \Delta\tau, \tag{55}$$

where $2\Delta\tau = j/i$ such that $j = 2\Delta\tau i$. Note that by definition, $\Delta\tau \in [0, \frac{1}{2}]$. Also, note that unless, $j = 0$ when $i = 0$, a singularity would appear. However, in purely symmetric systems, there is no singularity since both $j = 0$ and $i = 0$ when $V = 0$.

The transport number requires some modification for asymmetric salt concentrations and symmetric diffusion coefficients ($D_- = 1$) [33]. Namely, the subtraction of the already known $j_{i=0}$ from $j$ such that $2\Delta\tau = (j - j_{i=0})/i$ [or $j = 2\Delta\tau i + j_{i=0}$]. Here, too, $\Delta\tau \in [0, \frac{1}{2}]$. A similar modification is needed for the case of asymmetric salt concentrations and general $D_-$, we suggest a similar ansatz, $2\Delta\tau = (j - j_{i=0})/i$, which utilizes the still unknown $j_{i=0}$ {which is not to be confused with the already-known $\hat{j}_{i=0}$ [Eq. (46)]}. This expression can also be rewritten as

$$j = 2\Delta\tau i + j_{i=0}. \tag{56}$$

The substitution of Eqs. (36) and (56) into Eq. (47) leads to a complicated function $f(i, \Delta\tau; j_{i=0}) = 0$, which is not given here as it is exceedingly long and requires evaluation in a mathematical compiler (we have used Mathematica [51]). Following our previous works [33,50], we take the Taylor series of $f(i, \Delta\tau; j_{i=0}) = 0$ around $i = 0$ such that $f(i, \Delta\tau) \approx f(i=0) + (\partial f / \partial i)_{i=0} i$ and require that both leading order terms are zero. The first $f(i=0)$ term is independent of the transport number, and requiring that it is zero yields

$$j_{i=0} = \frac{2D_-}{(1+D_-)^2 L}\left[(1+D_-)\hat{j}_{i=0} L + (1-D_-)\Sigma_s \ln\left(\frac{Q_{\text{right}}}{Q_{\text{left}}}\right)\right], \tag{57}$$

where

$$\begin{aligned} Q_{\text{left}} &= \Sigma_s(1-D_-) + 2(1+D_-)S_{\text{left}} \\ Q_{\text{right}} &= \Sigma_s(1-D_-) + 2(1+D_-)S_{\text{right}} \end{aligned}. \tag{58}$$

Finally, requiring that $(\partial f / \partial i)_{i=0} = 0$ yields

$$\Delta\tau = \frac{(1-D_-)}{2(1+D_-)} + \frac{4D_-^2(D_- - 1)\Sigma_s^2}{(1+D_-)^2 Q_{\text{left}} Q_{\text{right}}} \frac{\hat{j}_{i=0}}{j_{i=0}} + \frac{4D_-^2 \Sigma_s}{j_{i=0} L(1+D_-)^3} \ln\left(\frac{Q_{\text{left}}}{Q_{\text{right}}}\right). \tag{59}$$

**D. Ohmic conductance, $g_{\text{Ohmic}}$, and zero-current voltage, $V_{i=0}$**

Inserting Eq. (36) into Eq. (54) yields the $i - V$ response

$$V = [(D_- - 1)i + (D_- + 1)j]\frac{L}{2D_-\Sigma_s} - \hat{j}_{\hat{i}=0}\frac{L}{\Sigma_s} + \hat{V}_{\hat{i}=0}. \tag{60}$$

Then, inserting Eq. (56) into Eq. (60) yields the $i-V$ near $i=0$

$$V = \frac{[D_- - 1 + 2(1+D_-)\Delta\tau]L}{2D_-\Sigma_s}i + \frac{(1+D_-)Lj_{i=0}}{2D_-\Sigma_s} - \hat{j}_{\hat{i}=0}\frac{L}{\Sigma_s} + \hat{V}_{\hat{i}=0}. \tag{61}$$

Equation (61) can be rewritten elegantly as a shifted Ohm's law

$$V = r_{\text{Ohmic}}i + V_{i=0}, \tag{62}$$

$$r_{\text{Ohmic}} = g_{\text{Ohmic}}^{-1} = \frac{[D_- - 1 + 2(1+D_-)\Delta\tau]}{2D_-}\frac{L}{\Sigma_s}, \tag{63}$$

$$V_{i=0} = \left[\frac{(1+D_-)j_{i=0}}{2D_-} - \hat{j}_{\hat{i}=0}\right]\frac{L}{\Sigma_s} + \hat{V}_{\hat{i}=0}, \tag{64}$$

where $r_{\text{Ohmic}} = g_{\text{Ohmic}}^{-1}$ is the Ohmic resistance multiplied by the area and reciprocal to the all-important Ohmic conductance (density), $g_{\text{Ohmic}}$, and $V_{i=0}$ is the *measurable* shifted voltage.

## V. RESULTS AND DISCUSSION

We divide our results section into three. First, we provide the dimensional forms of all the key results and their reduction in various limits (Sec. V.A). Second, we shall discuss the main results as we compare the various limits to non-approximated numerical simulations (Sec. V.B). Finally, we shall address the issues of calculating the zero-voltage current, $\tilde{i}_{\tilde{V}=0}$, and the power, $\tilde{P}$ (Sec. V.C).

### A. Dimensional form of key results

#### 1. General model

The final expressions for the currents and voltages depend on several key parameters, all of which need to be dimensionalized [using Eq. (10)]. If $\tilde{S}_k = S_k \tilde{c}_{bulk}$, then Eq. (44) is

$$\tilde{S}_{\text{left}} = \tfrac{1}{2}\sqrt{\tilde{\Sigma}_s^2 + 4\tilde{c}_{\text{left}}^2}, \quad \tilde{S}_{\text{right}} = \tfrac{1}{2}\sqrt{\tilde{\Sigma}_s^2 + 4\tilde{c}_{\text{right}}^2}. \tag{65}$$

In a similar manner, if $\tilde{Q}_k = Q_k \tilde{c}_{bulk}\tilde{D}_+$ then Eq. (58)

$$\begin{aligned}\tilde{Q}_{\text{left}} &= \tilde{\Sigma}_s(\tilde{D}_+ - \tilde{D}_-) + 2(\tilde{D}_+ + \tilde{D}_-)\tilde{S}_{\text{left}} \\ \tilde{Q}_{\text{right}} &= \tilde{\Sigma}_s(\tilde{D}_+ - \tilde{D}_-) + 2(\tilde{D}_+ + \tilde{D}_-)\tilde{S}_{\text{right}}\end{aligned}. \tag{66}$$

Then, the various fluxes [Eqs. (46) and (57)] and potentials [Eqs. (53) and (64)] can be dimensionalized with Eq. (10) yielding

$$\tilde{\tilde{j}}_{\hat{i}=0} = 2\tilde{D}_+(\tilde{S}_{\text{left}} - \tilde{S}_{\text{right}})/\tilde{L}, \tag{67}$$

$$\tilde{j}_{i=0} = \frac{2\tilde{D}_-}{(\tilde{D}_+ + \tilde{D}_-)}\tilde{\tilde{j}}_{\hat{i}=0} + \frac{2\tilde{D}_-\tilde{D}_+(\tilde{D}_+ - \tilde{D}_-)}{(\tilde{D}_+ + \tilde{D}_-)^2 \tilde{L}}\tilde{\Sigma}_s \ln\left(\frac{\tilde{Q}_{\text{right}}}{\tilde{Q}_{\text{left}}}\right), \tag{68}$$

$$\tilde{\tilde{V}}_{\hat{i}=0} = \tilde{V}_{th}\left[\ln\left(\frac{\tilde{c}_{\text{right}}}{\tilde{c}_{\text{left}}}\right) - \ln\left(\frac{\tilde{S}_{\text{right}} + \frac{1}{2}\tilde{\Sigma}_s}{\tilde{S}_{\text{left}} + \frac{1}{2}\tilde{\Sigma}_s}\right)\right], \tag{69}$$

$$\tilde{V}_{i=0} = \left[\frac{(\tilde{D}_+ + \tilde{D}_-)\tilde{j}_{i=0}}{2\tilde{D}_-} - \tilde{\tilde{j}}_{\hat{i}=0}\right]\frac{\tilde{L}\tilde{V}_{th}}{\tilde{D}_+\tilde{\Sigma}_s} + \tilde{\tilde{V}}_{\hat{i}=0}. \tag{70}$$

The transport number remains non-dimensional ($\Delta\tau = \Delta\tau$) and is given by

$$\Delta\tau = \frac{(\tilde{D}_+ - \tilde{D}_-)}{2(\tilde{D}_+ + \tilde{D}_-)} + \frac{4\tilde{D}_-^2\tilde{D}_+(\tilde{D}_- - \tilde{D}_+)\tilde{\Sigma}_s^2}{(\tilde{D}_+ + \tilde{D}_-)^2\tilde{Q}_{\text{left}}\tilde{Q}_{\text{right}}}\frac{\tilde{\tilde{j}}_{\hat{i}=0}}{\tilde{j}_{i=0}} + \frac{4\tilde{D}_+^2\tilde{D}_-^2\tilde{\Sigma}_s}{\tilde{j}_{i=0}\tilde{L}(\tilde{D}_+ + \tilde{D}_-)^3}\ln\left(\frac{\tilde{Q}_{\text{left}}}{\tilde{Q}_{\text{right}}}\right). \tag{71}$$

The Ohmic conductance is dimensionalized by $\tilde{F}\tilde{j}_0/\tilde{V}_{th}$

$$\tilde{g}_{\text{Ohmic}} = \frac{\tilde{D}_+\tilde{F}^2}{\tilde{R}\tilde{T}}\frac{2\tilde{D}_-\tilde{\Sigma}_s}{\tilde{D}_- - \tilde{D}_+ + 2(\tilde{D}_+ + \tilde{D}_-)\Delta\tau}\frac{1}{\tilde{L}}. \tag{72}$$

Transfomation of all flux, currents, and conductance densities terms to geometry-dependent terms is via in Eq. (1) such that $\tilde{J}_\pm = \tilde{j}_\pm\tilde{A}, \tilde{J} = \tilde{j}\tilde{A}, \tilde{I} = \tilde{i}\tilde{A}$. We shall continue to focus on the geometry-independent terms (i.e., the densities).

Since Eqs.(65)-(72) are new and have yet to be investigated, we will first demonstrate that we are able to recapitulate three known limiting models [equal diffusion coefficients (Sec. V.A.2), symmetric salt concentrations (Sec V.A.3), and symmetric salt concentrations with equal diffusion coefficients (Sec V.A.4)]. Then, we will analyze the each of these scenarios in Sec. V.B to better understand each limit. Finally, we will analyze the new model.

### 2. Equal diffusion coefficients ($\tilde{D}_+ = \tilde{D}_- = \tilde{D}$)

We shall now reduce Eqs.(65)-(72), for the case of $\tilde{D}_+ = \tilde{D}_- = \tilde{D}$. The expressions for $\tilde{S}_{\text{left}}$ and $\tilde{S}_{\text{right}}$ remain unchanged, while $\tilde{Q}_{\text{left}} = 4\tilde{D}\tilde{S}_{\text{left}}$ and $\tilde{Q}_{\text{right}} = 4\tilde{D}\tilde{S}_{\text{right}}$. We also find that

$\tilde{j}_{i=0} \equiv \tilde{\hat{j}}_{\hat{i}=0}$ and $\tilde{V}_{i=0} \equiv \tilde{\hat{V}}_{\hat{i}=0}$ [Eq. (69)]. The transport number and Ohmic conductance are given by

$$\Delta \tau(\tilde{D}_+ = \tilde{D}_-) = \frac{\tilde{D}\tilde{\Sigma}_s}{2\tilde{\hat{j}}_{\hat{i}=0}\tilde{L}} \ln \frac{\tilde{S}_{\text{left}}}{\tilde{S}_{\text{right}}}, \tag{73}$$

$$\tilde{g}_{\text{Ohmic}}(\tilde{D}_+ = \tilde{D}_-) = 2\frac{\tilde{D}\tilde{F}^2}{\tilde{R}\tilde{T}} \frac{\tilde{S}_{\text{left}} - \tilde{S}_{\text{right}}}{\ln(\tilde{S}_{\text{left}}/\tilde{S}_{\text{right}})} \frac{1}{\tilde{L}}. \tag{74}$$

All these results are consistent with our past work [33]. Note the similarity of Eq. (74) to (27) where $\tilde{S}_{\text{left}}$ and $\tilde{S}_{\text{right}}$ now play the more general role of $C_{\text{left}}$ and $C_{\text{right}}$.

### 3. No concentration gradients ($\tilde{c}_{\text{left}} = \tilde{c}_{\text{right}} = \tilde{c}_{\text{bulk}}$)

In a similar manner, we can reduce Eqs.(65)-(72), for the case of $\tilde{c}_{\text{left}} = \tilde{c}_{\text{right}} = \tilde{c}_{\text{bulk}}$. Unsurprisingly, we find that $\tilde{S}_{\text{left}} = \tilde{S}_{\text{right}} = \tilde{S}_{\text{bulk}}(\tilde{c}_{\text{bulk}})$ and $\tilde{Q}_{\text{left}} = \tilde{Q}_{\text{right}} = \tilde{Q}_{\text{bulk}}(\tilde{c}_{\text{bulk}})$. Also, trivially, $\tilde{j}_{i=0} = \tilde{\hat{j}}_{\hat{i}=0} = 0$ and $\tilde{V}_{i=0} = \tilde{\hat{V}}_{\hat{i}=0} = 0$. The transport number and Ohmic conductance are

$$\Delta \tau(\tilde{c}_{\text{left}} = \tilde{c}_{\text{right}} = \tilde{c}_{\text{bulk}}) = \frac{2(\tilde{D}_+ - \tilde{D}_-)\tilde{S}_{\text{bulk}} + (\tilde{D}_+ + \tilde{D}_-)\tilde{\Sigma}_s}{4(\tilde{D}_+ + \tilde{D}_-)\tilde{S}_{\text{bulk}} + 2(\tilde{D}_+ - \tilde{D}_-)\tilde{\Sigma}_s}, \tag{75}$$

$$\tilde{g}_{\text{Ohmic}}(\tilde{c}_{\text{left}} = \tilde{c}_{\text{right}} = \tilde{c}_{\text{bulk}}) = \frac{\tilde{F}^2}{2\tilde{R}\tilde{T}}[(\tilde{D}_+ - \tilde{D}_-)\tilde{\Sigma}_s + 2_+(\tilde{D}_- + \tilde{D}_+)\tilde{S}_{\text{bulk}}]\frac{1}{\tilde{L}} = \frac{\tilde{F}^2 \tilde{Q}_{\text{bulk}}}{2\tilde{R}\tilde{T}\tilde{L}}. \tag{76}$$

A similar expression was cleverly suggested in the work of Noh and Aluru [52]. However, to the best of our knowledge, and based on our discussion with Prof. Aluru, a rigorous derivation of this expression has still to be given.

### 4. No concentration gradients for symmetric diffusion coefficients ($\tilde{c}_{\text{left}} = \tilde{c}_{\text{right}} = \tilde{c}_{\text{bulk}}$ and $\tilde{D}_+ = \tilde{D}_- = \tilde{D}$)

When taking the limit of $\tilde{c}_{\text{left}} = \tilde{c}_{\text{right}} = \tilde{c}_{\text{bulk}}$ for the results of Sec. V.A.2 or the limit of $\tilde{D}_+ = \tilde{D}_- = \tilde{D}$ for the results of Sec V.A.3, the same result is recapitulated. Namely, $\tilde{S}_{\text{left}} = \tilde{S}_{\text{right}} = \tilde{S}_{\text{bulk}}(\tilde{c}_{\text{bulk}})$ and $\tilde{Q}_{\text{left}} = \tilde{Q}_{\text{right}} = \tilde{Q}_{\text{bulk}}(\tilde{c}_{\text{bulk}}) = 4\tilde{S}_{\text{bulk}}(\tilde{c}_{\text{bulk}})$, which in turn yields that $\tilde{j}_{i=0} = \tilde{\hat{j}}_{\hat{i}=0} = 0$ and $\tilde{V}_{i=0} = \tilde{\hat{V}}_{\hat{i}=0} = 0$. Thereafter, we recover the transport number

$$\Delta \tau(\tilde{D}_+ = \tilde{D}_-, c_{\text{left}} = c_{\text{right}} = c_{\text{bulk}}) = \tilde{\Sigma}_s / (4\tilde{S}_{\text{bulk}}), \tag{77}$$

which yields the famous square-root law for the conductance

$$\tilde{g}_{\text{Ohmic}}(\tilde{D}_+ = \tilde{D}_-, \tilde{c}_{\text{left}} = \tilde{c}_{\text{right}} = \tilde{c}_{\text{bulk}}) = 2\frac{\tilde{D}\tilde{F}^2}{\tilde{R}\tilde{T}}\frac{\tilde{S}_{\text{bulk}}}{\tilde{L}} = \frac{\tilde{D}\tilde{F}^2 \tilde{c}_{\text{bulk}}}{\tilde{R}\tilde{T}}\sqrt{\left(\frac{\tilde{\Sigma}_s}{\tilde{c}_{\text{bulk}}}\right)^2 + 4}\frac{1}{\tilde{L}}, \qquad (78)$$

which rationalize (and follow) the pioneering experiments of Stein et al. [45].

### B. Analysis of conductance and zero-current voltage

Before proceeding with the analysis of the general solution, it is worthwhile to work our way back. In Sec. V.B.1, we will start with a short discussion on the simplest and most investigated scenario of symmetric salt concentrations with equal diffusion coefficients (Sec V.A.4). We will then leverage this solution and that of the symmetric salt concentrations scenario (Sec V.A.3) to discuss a common misconception in the literature – the superposition approach. In Sec. V.B.2, we will briefly discuss the equal diffusion coefficient with asymmetric salt concentrations (Sec. V.A.2). Thereafter, in Sec. V.B.3, we will turn to the general scenario of systems with asymmetric diffusion coefficients subject to asymmetric salt concentrations (Sec V.A.1) – this general scenario serves as the replacement theory for GHK.

Each of the scenarios below will be compared to the respective non-approximated numerical simulations. In all scenarios, it will be observed that we have remarkable correspondence between theory and simulations (which will always be given as purple squares). For brevity, unless needed, we will not repeat this otherwise, obvious observation.

#### 1. Symmetric salt concentrations ($\tilde{c}_{\text{left}} = \tilde{c}_{\text{right}}$)

Equation (78)

$$\tilde{g}_{\text{Ohmic}} = \frac{\tilde{D}\tilde{F}^2 \tilde{c}_{\text{bulk}}}{\tilde{R}\tilde{T}}\sqrt{\left(\frac{\tilde{\Sigma}_s}{\tilde{c}_{\text{bulk}}}\right)^2 + 4}\frac{1}{\tilde{L}},$$

is controlled by a single control parameter – the non-dimensional excess counterion concentration $\Sigma_s = \tilde{\Sigma}_s / \tilde{c}_{\text{bulk}}$. In the limit that $\tilde{\Sigma}_s \ll \tilde{c}_{\text{bulk}}$, the conductance is determined by the bulk conductance such that

$$\tilde{g}_{\text{Ohmic}}(\tilde{\Sigma}_s \ll \tilde{c}_{\text{bulk}}) = \frac{2\tilde{D}\tilde{F}^2 \tilde{c}_{\text{bulk}}}{\tilde{R}\tilde{T}\tilde{L}}. \qquad (79)$$

This is the part of the curve with a slope of 1, previously shown in Figure 3 and now shown again in Figure 6. In the opposite limit, $\tilde{\Sigma}_s \gg \tilde{c}_{\text{bulk}}$, the conductance is determined by the excess counterion concentration such that

$$\tilde{g}_{\text{Ohmic}}(\tilde{N} \gg \tilde{c}_{\text{bulk}}) = \frac{\tilde{D}\tilde{F}^2\tilde{\Sigma}_s}{\tilde{R}\tilde{T}\tilde{L}}, \tag{80}$$

is explicitly independent of the concentration – this is the part of the curve with a slope of 0.

In the literature, one can often find the ohmic conductance given as the superposition of these two limits [sum of Eq. (79) and (80)]

$$\tilde{g}_{\text{Ohmic-superposition}} = \tilde{g}_{\text{Ohmic}}(\tilde{\Sigma}_s \gg \tilde{c}_{\text{bulk}}) + \tilde{g}_{\text{Ohmic}}(\tilde{\Sigma}_s \ll \tilde{c}_{\text{bulk}}) = \frac{\tilde{D}\tilde{F}^2\tilde{c}_{\text{bulk}}}{\tilde{R}\tilde{T}}\left(\frac{\tilde{\Sigma}_s}{\tilde{c}_{\text{bulk}}} + 2\right)\frac{1}{\tilde{L}}. \tag{81}$$

Naturally, based on how this empirical approximation was constructed, it matches Eq. (78) perfectly in the two limits of $\tilde{\Sigma}_s \ll \tilde{c}_{\text{bulk}}$ and $\tilde{\Sigma}_s \gg \tilde{c}_{\text{bulk}}$. However, both models make different predictions for the region $\tilde{\Sigma}_s = \tilde{c}_{\text{bulk}}$. Figure 6 compares Eqs. (78) and (81) to exact numerical simulations from which it is evident that there is perfect correspondence to Eq. (78) for all values of $\tilde{\Sigma}_s/\tilde{c}_{\text{bulk}}$. In contrast, Eq. (81) has good correspondence only at the two limits it was constructed to satisfy. Similar to how the concentrations predicted by GHK were not vastly different than those predicted by the correct model, here, too, the differences between the two models for $\tilde{\Sigma}_s = \tilde{c}_{\text{bulk}}$ are are not vastly different, except in what they represent. Equation (81) represents a phenomenological prediction for the conductance and has been designed to be simple and appealing. In contrast, Eq. (78) has been derived without any empirical assumption and is the limiting conductance of three different models (two shown here and another one discussed briefly now). In Ref. [50], the total resistance of a "realistic " system was derived. Realistic refers to the fact that the combined effects of the microchannels and the nanochannel are accounted for. It was shown that when the effects of the microchannel resistances are neglected, Eq. (78) is reproduced. See Ref. [50] for a detailed discussion and thorough derivation regarding the effects of the microchannel or Ref. [13] for a detailed discussion of the same model without the involved mathematics).

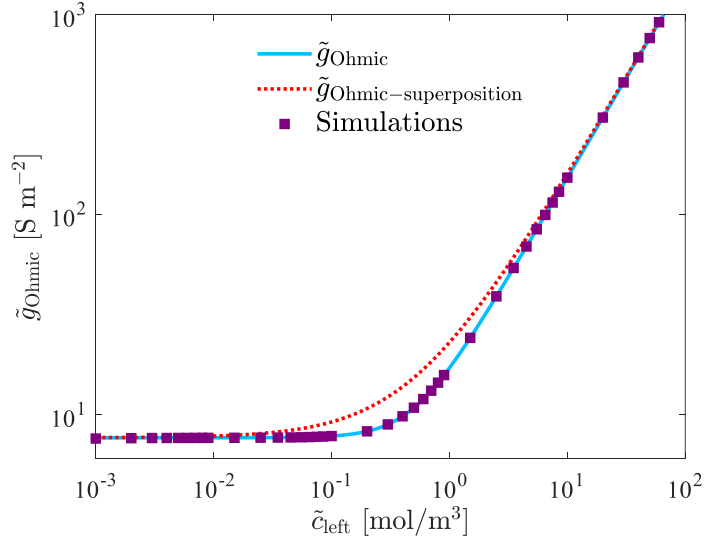

**Figure 6**. Conductance versus concentration plot for KCl electrolyte [i.e., symmetric diffusion coefficients ($\tilde{D}_K = \tilde{D}_{Cl}$)] for symmetric concentrations ($\tilde{c}_{left} = \tilde{c}_{right} = \tilde{c}_{bulk}$) comparing the correct square-root law, $\tilde{g}_{Ohmic}$ [Eq. (78)], and the common but incorrect empirical superposition law, $\tilde{g}_{Ohmic\text{-}superposition}$ [Eq. (81)] to numerical simulations. Simulation parameters are detailed in Table 2.

To further highlight this issue, we will now compare Eq. (76) to a derivative model of Eq. (81), namely the superposition approach for ions with different diffusion coefficients

$$\tilde{g}_{Ohmic\text{-}superposition} = \frac{\tilde{F}^2}{\tilde{R}\tilde{T}}[\tilde{D}_+\tilde{\Sigma}_s + (\tilde{D}_+ + \tilde{D}_-)\tilde{c}_{bulk}]\frac{1}{\tilde{L}}. \tag{82}$$

Note that the first term here still represents the excess counterion contribution, whereas the second term represents the bulk contribution of an asymmetric salt. Figure 7 compares numerical simulations to Eq. (76) and Eq. (82). Unsurprisingly, we find excellent correspondence to Eq. (76) and that Eq. (82) fails for the same reason Eq. (81) failed. Importantly, we can see that most of the behavior/trends of Eq. (78) also holds for Eq. (76). Namely, at high concentrations, the slope is 1, and at low concentrations, the slope is still 0. Here, the low concentration value of the conductance of both Eqs. (76) and (78) are different as the diffusion coefficients of KCl ($\tilde{D}_+ = \tilde{D}_K$) and HCl ($\tilde{D}_+ = \tilde{D}_H$) differ.

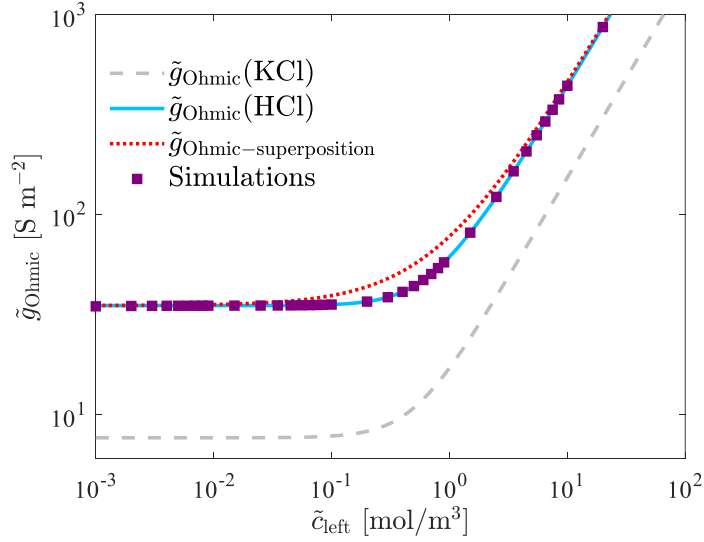

**Figure 7**. Conductance versus concentration plot for HCl with symmetric bulk concentrations ($\tilde{c}_{\text{left}} = \tilde{c}_{\text{right}} = \tilde{c}_{\text{bulk}}$) comparing the conductance, $\tilde{g}_{\text{Ohmic}}$ [Eq. (76)], and the common but incorrect empirical superposition law for arbitrary diffusion coefficients, $\tilde{g}_{\text{Ohmic-superposition}}$ [Eq. (82)] to numerical simulations. For the sake of comparison, we have kept $\tilde{g}_{\text{Ohmic}}$ for KCl in grey [Eq. (78)]. Simulation parameters are detailed in Table 2.

We will thus argue, as we have in the past [13,33,50,53], that the superposition approach of adding conductances (and its derivative models) should be discontinued, and a more rigorous approach should be adopted.

### 2. Asymmetric salt concentrations ($\tilde{c}_{\text{left}} \neq \tilde{c}_{\text{right}}$) with symmetric diffusion coefficients ($\tilde{D}_+ = \tilde{D}_-$)

The results of the equal diffusion coefficient with asymmetric salt concentrations (Sec. V.A.2) were covered thoroughly in our recent work [33]. Thus, we provide only a brief discussion.

Figure 8(a) shows $\tilde{g}_{\text{Ohmic}}$ [Eq. (74)] for three ratios of $\tilde{c}_{\text{right}}/\tilde{c}_{\text{left}}$ versus $\tilde{c}_{\text{left}}$. It can be observed that at low concentrations, the conductance saturates to a value determined by the surface conductance [Eq.(80)]. At high concentrations, all the curves are parallel to the curve of KCL and have a slope of 1. At the limit of $\tilde{\Sigma}_s \ll \tilde{c}_{\text{bulk}}$ (or when $\tilde{\Sigma}_s/\tilde{c}_{\text{bulk}} \to 0$), Eq. (74) can be written as

$$\tilde{g}_{\text{Ohmic}}(\tilde{N} \ll \tilde{c}_{\text{bulk}}) = 2\frac{\tilde{D}\tilde{F}^2}{\tilde{R}\tilde{T}}\frac{\tilde{c}_{\text{left}} - \tilde{c}_{\text{right}}}{\ln(\tilde{c}_{\text{left}}/\tilde{c}_{\text{right}})}\frac{1}{\tilde{L}}. \qquad (83)$$

This equation is the dimensional conductance calculated in Eq. (27).

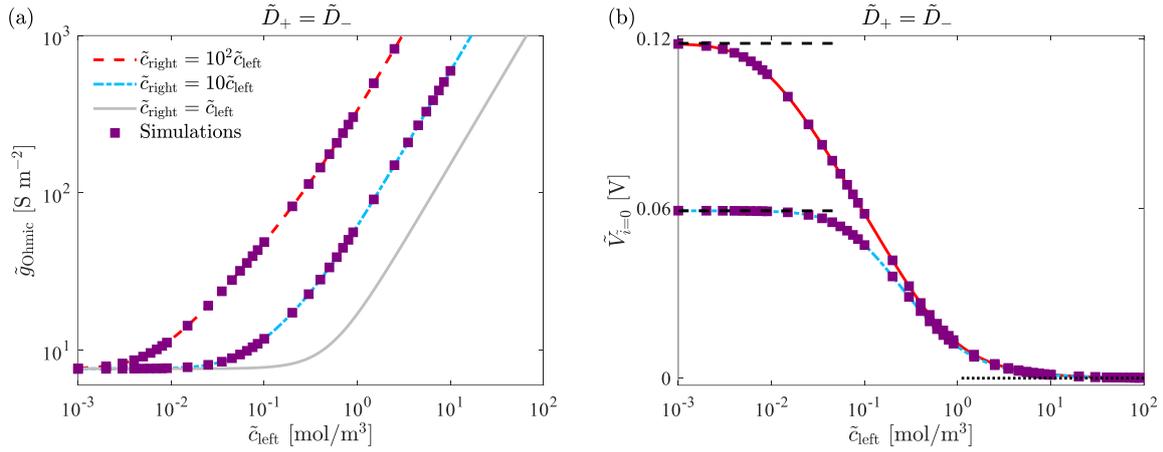

**Figure 8**. (a) Conductance versus concentration plot for a KCl electrolyte for non-symmetric concentrations ($\tilde{c}_{\text{left}} \neq \tilde{c}_{\text{right}}$) comparing $\tilde{g}_{\text{Ohmic}}$ [Eq. (74)] to numerical simulations. For the sake of comparison, we have kept the conductance for a symmetric salt concentration ($\tilde{c}_{\text{left}} = \tilde{c}_{\text{right}} = \tilde{c}_{\text{bulk}}$) [Eq. (78)]. (b) The zero current voltage, $\tilde{V}_{i=0}$ [$\equiv \tilde{\hat{V}}_{\hat{i}=0}$, Eq. (69)], versus the concentration. The Nernst potential [$\tilde{V}_{\text{Nernst}} = \tilde{V}_{th} \ln(\tilde{c}_{\text{right}} / \tilde{c}_{\text{left}})$, Eq. (3)] is denoted by the dashed black line while $\tilde{V}_{\text{Goldman}}$ [Eq. (4)] is denoted by the dotted black line. Simulation parameters are detailed in Table 2.

Finally, we note that as we increase the concentration ratio $\tilde{c}_{\text{right}} / \tilde{c}_{\text{left}}$, we see that the transition from being bulk-dominated ($\tilde{\Sigma}_s \ll \tilde{c}_{\text{bulk}}$) to $\tilde{\Sigma}_s$-dominated ($\tilde{\Sigma}_s \gg \tilde{c}_{\text{bulk}}$) occurs at lower concentrations. This can be expected as, here, the concentration on the right is higher than on the left, and hence, lower concentrations are needed to become highly selective. In contrast, if the concentration ratio were smaller than one, the transition point would shift to higher concentrations (not shown here; see Ref. [33]).

Figure 8(b) shows the zero-current voltage $\tilde{V}_{i=0}$ [$\equiv \tilde{\hat{V}}_{\hat{i}=0}$, Eq. (69)]. At low concentrations, the value saturates to the Nernst equation [Eq. (3)]. At high concentrations, the value saturates to the correct expression derived by Goldman [Eq. (4)], which predicts this value to be zero. Importantly, our model is able to capture the behavior of $\tilde{V}_{i=0}$ for all values of $\tilde{c}_{\text{right}}$ and $\tilde{c}_{\text{left}}$. In a similar manner to the behavior of the conductance, we see that as we increase the concentration ratio $\tilde{c}_{\text{right}} / \tilde{c}_{\text{left}}$, the concentration that reaches the Nernst potential [Eq. (3)] occurs at a lower concentration. This, too, makes sense, as a lower concentration is needed to achieve ideal selectivity.

## 3. Asymmetric salt concentrations ($\tilde{c}_{\text{left}} \neq \tilde{c}_{\text{right}}$) with asymmetric diffusion coefficients ($\tilde{D}_+ \neq \tilde{D}_-$)

Figure 9 shows the behavior of $\tilde{g}_{\text{Ohmic}}$ for two different values of $\tilde{c}_{\text{right}} / \tilde{c}_{\text{left}}$ for several electrolytes (LiCl, NaCl, KCl, HCl). All of the trends of the KCl electrolyte discussed in the previous paragraph still hold. The one major difference that can be observed is that as $\tilde{D}_+$ is increased, the conductance increases as expected. This can be expected since the low-concentration conductance is linear with $\tilde{D}_+$.

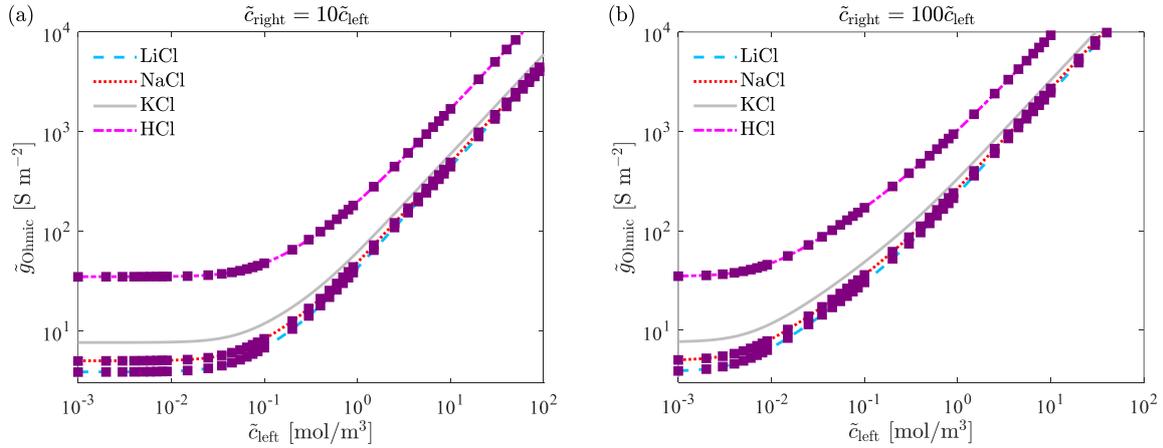

**Figure 9**. Conductance versus concentration plot for several electrolytes (LiCl, NaCl, KCl, HCl) for (a) $\tilde{c}_{\text{right}} = 10\tilde{c}_{\text{left}}$ and (b) $\tilde{c}_{\text{right}} = 10^2 \tilde{c}_{\text{left}}$. We compare the general $\tilde{g}_{\text{Ohmic}}$ [Eq. (72)] to numerical simulations. For the sake of comparison, we have kept $\tilde{g}_{\text{Ohmic}}$ for KCl [Eq. (74), previously shown in Figure 8(a)]. Simulation parameters are detailed in Table 2.

Figure 10 shows the behavior of $\tilde{V}_{\tilde{i}=0}$ for the same scenarios shown in Figure 9. The most important observation is that $\tilde{V}_{\tilde{i}=0}$ at low concentrations is always the Nernst potential, which is independent of the diffusion coefficients, while at high concentrations, $\tilde{V}_{\tilde{i}=0}$ reduces to $\tilde{V}_{\text{Goldman}}$ [Eq. (4)] which predicted the change in the sign of the value depending on whether $\tilde{D}_+$ is bigger or smaller than $\tilde{D}_-$. In particular, it can be observed that when $\tilde{D}_+ \leq \tilde{D}_-$, $\tilde{V}_{\tilde{i}=0}$ changes sign (positive to negative, or vice versa depending on the ratio $\tilde{c}_{\text{right}} / \tilde{c}_{\text{left}}$) such that there is a concentration where $\tilde{V}_{\tilde{i}=0} = 0$ and the $\tilde{i} - \tilde{V}$ crosses the origin leading to both $\tilde{i}_{\tilde{V}=0} = 0$ and a zero power output. This surprising result will soon be demonstrated.

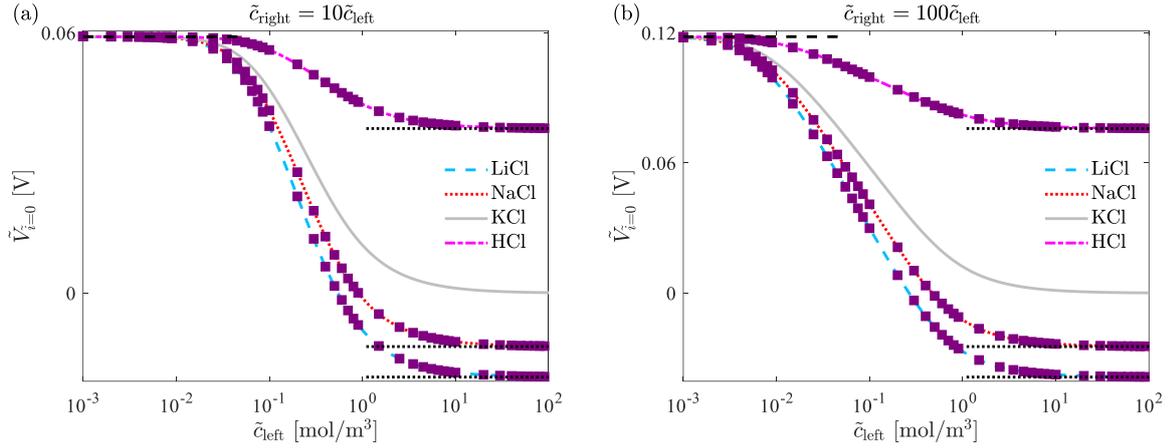

**Figure 10**. The zero current voltage, $\tilde{V}_{\tilde{i}=0}$ versus concentration for several electrolytes (LiCl, NaCl, KCl, HCl) for (a) $\tilde{c}_{\text{right}} = 10\tilde{c}_{\text{left}}$ and (b) $\tilde{c}_{\text{right}} = 10^2 \tilde{c}_{\text{left}}$. We compare the general $\tilde{V}_{\tilde{i}=0}$ [Eq. (70)] to numerical simulations. For the sake of comparison, we have kept $\tilde{V}_{\tilde{i}=0}$ for KCl [Eq. (69)], previously shown in Figure 8(b). The Nernst potential [Eq. (3)] is denoted by the dashed black line while $\tilde{V}_{\text{Goldman}}$ [Eq. (4)] is denoted by the dotted black line. Simulation parameters are detailed in Table 2.

One last comment on the combined behavior of $\tilde{g}_{\text{Ohmic}}$ (Figure 9) and $\tilde{V}_{\tilde{i}=0}$ (Figure 10) also leads to another prediction, which will shortly be verified. Consider the limit of high concentrations when $\tilde{V}_{\tilde{i}=0}$ saturates to the constant $\tilde{V}_{\text{Goldman}}$ (which, except for KCl, is not zero) and that conductance continues to increase whereby $\tilde{g}_{\text{Ohmic}} \sim \tilde{c}_{\text{left}}$. If $\tilde{V}_{\tilde{i}=0}$ is constant, but the conductance increases, the second intersection $\tilde{i}_{\tilde{V}=0}$ must also increase (in the absolute-value sense). Therefore, we predict (and shortly demonstrate this), except for KCl, that at very high concentrations, $\tilde{i}_{\tilde{V}=0} \sim \tilde{c}_{\text{left}}$.

## C. Zero-voltage current, $\tilde{i}_{\tilde{V}=0}$, and power, $\tilde{P}$

To calculate the available power density from RED, $\tilde{P}$, we need to calculate the zero-voltage current, $\tilde{i}_{\tilde{V}=0}$, and the differential conductance (or differential resistance), $\tilde{g}_{\tilde{V}=0} = (d\tilde{i}/d\tilde{V})_{\tilde{V}=0}$. Before proceeding with the calculations, it is worthwhile to demonstrate that our previous statement that $\tilde{g}_{\tilde{i}=0} \neq \tilde{g}_{\tilde{V}=0}$ is correct (note that this already is illustrated in Figure 2).

To that end, we return to the most general form of the resultant solution – these are Eqs. (47) and (54), given here again in non-dimensional form and in terms of the "reduced" fluxes

$$f(\hat{i}, \hat{j}) = \frac{\hat{i}}{\hat{j}} \Sigma_s \ln\left(\frac{\hat{i}\Sigma_s - 2\hat{j}S_{\text{right}}}{\hat{i}\Sigma_s - 2\hat{j}S_{\text{left}}}\right) + (\hat{j} - \hat{j}_{\tilde{i}=0})L = 0,$$

$$V = (\hat{j} - \hat{j}_{i=0})\frac{L}{\Sigma_s} + \hat{V}_{i=0}.$$

Consider the differential resistance (which is reciprocal to the differential conductance) that can be calculated using the chain rule

$$\frac{dV}{d\hat{i}} = \frac{dV}{d\hat{j}}\frac{d\hat{j}}{d\hat{i}} = \frac{L}{\Sigma_s}\frac{d\hat{j}}{d\hat{i}}. \tag{84}$$

$dV/d\hat{i}$ will be a constant only if $d\hat{j}/d\hat{i} = const$. Since there is a nonlinear dependence between $\hat{j}$ and $\hat{i}$ ensuring that $di/dj \neq const$, the two slopes $(d\hat{j}/d\hat{i})_{i=0}$ and $(d\hat{j}/d\hat{i})_{V=0}$ are different, yielding non-identical differential resistances.

We turn to finding $\tilde{i}_{\tilde{V}=0}$. When the voltage is zero, the "reduced" zero-voltage salt current density is given by

$$\hat{j}_{V=0} = \hat{j}_{i=0} - \frac{\Sigma_s}{L}\hat{V}_{i=0}, \tag{85}$$

(which in dimensional form is $\tilde{\hat{j}}_{V=0} = \tilde{\hat{j}}_{i=0} + \tilde{\Sigma}_s \tilde{D}_+ \tilde{\hat{V}}_{i=0}/\tilde{L}$). Inserting Eq. (85) into $f(\hat{i}, \hat{j}_{V=0}) = 0$ yields a transcendental equation for $\hat{i}_{\tilde{V}=0}$ that can be solved numerically. We solve this equation using the Newton-Raphson method and then transform the calculated $\hat{i} - \hat{j}$ terms to $i - j$ [using the inverse transformation of Eq. (36)]. These results are then (re-)dimensionalized to yield $\tilde{i} - \tilde{j}$.

Figure 11 presents the numerically calculated $\tilde{i}_{\tilde{V}=0}$. As can be expected, since $\tilde{V}_{i=0}$ is positive at low concentrations for all the different salts [Figure 10(a)], then $\tilde{i}_{\tilde{V}=0}$ must be negative at low concentrations. However, the high-concentration behavior is more complicated. Figure 10(a) shows that at high concentrations, $\tilde{V}_{i=0}$ for HCl is always positive. Similarly, the $\tilde{i}_{\tilde{V}=0}$ for HCl does not exhibit a change in sign (Figure 11). For KCl, it can also be observed that $\tilde{V}_{i=0}$ approaches zero from the positive value for KCl, such that there is no change of sign for $\tilde{V}_{i=0}$. Thus, $\tilde{i}_{\tilde{V}=0}$ does not exhibit a change in the sign. However, since both NaCl and LiCl exhibit a change in the sign of $\tilde{V}_{i=0}$ at high concentrations (relative to the low concentration values), then their $\tilde{i}_{\tilde{V}=0}$ must also exhibit a change of sign. This change of sign can be observed in Figure 11(a) [or the cusps in Figure 11(b)]. Also, we can observe that except for KCl, $\tilde{i}_{\tilde{V}=0} \sim \tilde{c}_{\text{left}}$. For KCl, we see that $\tilde{i}_{\tilde{V}=0}(\text{KCl})$ saturates to a constant value – this is because as $\tilde{V}_{\tilde{i}=0} \to 0$ and $\tilde{g}_{\text{Ohmic}} \sim \tilde{c}_{\text{left}}$, such that $\tilde{i}_{\tilde{V}=0}(\text{KCl}) \sim const$.

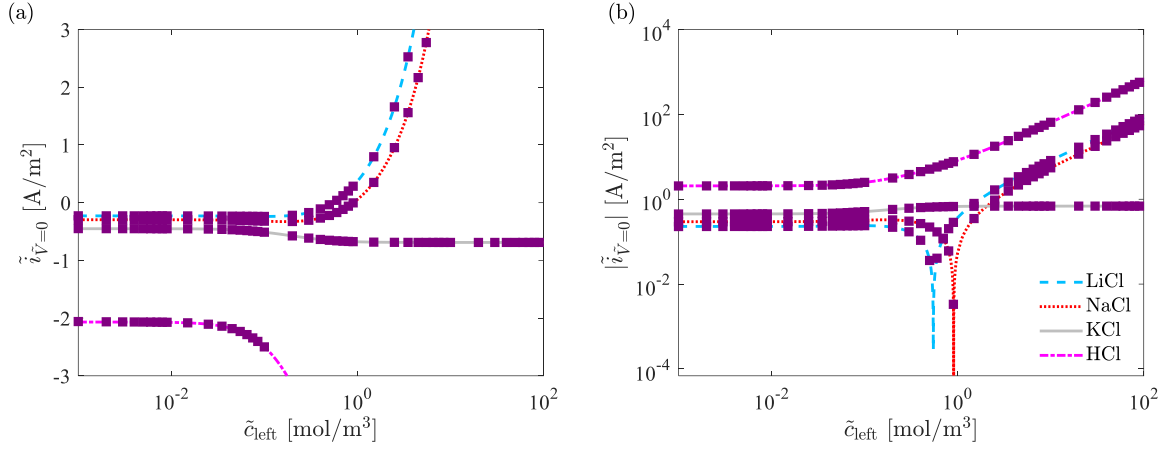

**Figure 11**. The zero-voltage current, $\tilde{i}_{\tilde{V}=0}$ versus the concentration for several electrolytes (LiCl, NaCl, KCl, HCl) for $\tilde{c}_{\text{right}} = 10\tilde{c}_{\text{left}}$ in (a) semi-log$_{10}$ scale and the absolute value (b) log$_{10}$-log$_{10}$ scale. The cusps in (b) for LiCl and NaCl are the crossover points from negative currents to positive – this crossover is evident in (a). Simulation parameters are detailed in Table 2.

To calculate the differential resistance, the above procedure is repeated for $\hat{j} = V\Sigma_s / L + \hat{j}_{\tilde{V}=0}$ for several values of $V$ in the vicinity of $V = 0$. Then, one can calculate $\tilde{g}_{\tilde{V}=0} = (d\tilde{i}/d\tilde{V})_{\tilde{V}=0}$. The difference between the two different conductances ($\tilde{g}_{\tilde{i}=0}$ and $\tilde{g}_{\tilde{V}=0}$) is rather small (even on a log$_{10}$-log$_{10}$ scale. To better distinguish between the two, we plot the ratio $\tilde{g}_{\tilde{i}=0}/\tilde{g}_{\tilde{V}=0}$ for two different concentrations $\tilde{c}_{\text{right}}/\tilde{c}_{\text{left}} = 10$ and $10^2$. Figure 12 shows that as the concentration ratio is increased, the difference between the two conductances becomes larger, the concentration region in which the changes are important shifts to the left, and the width of each curve (say the full width at half maximum) becomes larger. All these changes indicate that if one uses $\tilde{g}_{\tilde{i}=0}$ to estimate the power, depending on the concentration, the error is substantial.

Finally, we turn to calculate the highly desirable power density, $\tilde{P}_{\tilde{V}=0} = \tilde{i}_{\tilde{V}=0}^2 \tilde{g}_{\tilde{V}=0}^{-1}$, which is merely the multiplication of two quantities that have already been (numerically) calculated and discussed. Their multiplication, shown in Figure 13, yields some interesting results. We shall first address the general results that hold for all salts and then consider each salt separately.

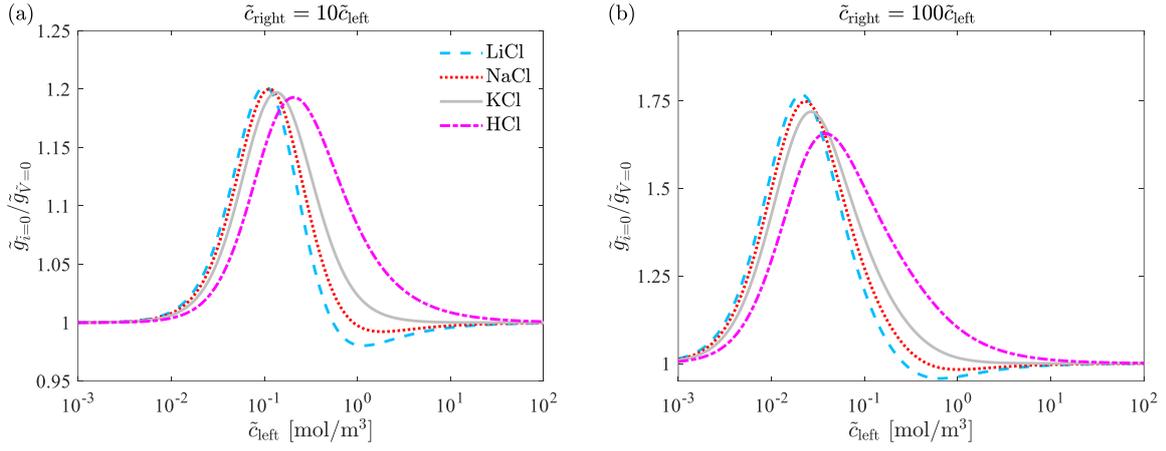

**Figure 12**. The ratio of $\tilde{g}_{\tilde{i}=0}/\tilde{g}_{\tilde{V}=0}$ versus the concentration for several electrolytes (LiCl, NaCl, KCl, HCl) for (a) $\tilde{c}_{\text{right}} = 10\tilde{c}_{\text{left}}$ and (b) $\tilde{c}_{\text{right}} = 10^2 \tilde{c}_{\text{left}}$ in a semi-log$_{10}$ scale. The parameters used here match those in Figure 11.

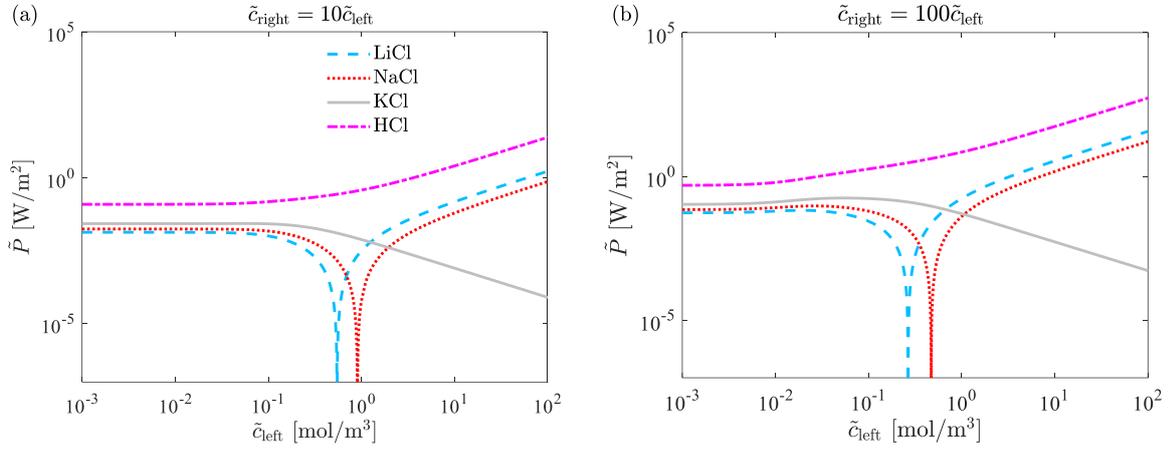

**Figure 13**. Power density, $\tilde{P}_{\tilde{V}=0} = \tilde{i}^2_{\tilde{V}=0} \tilde{g}^{-1}_{\tilde{V}=0}$, versus the concentration for several electrolytes (LiCl, NaCl, KCl, HCl) for (a) $\tilde{c}_{\text{right}} = 10\tilde{c}_{\text{left}}$ and (b) $\tilde{c}_{\text{right}} = 10^2 \tilde{c}_{\text{left}}$ in a log$_{10}$-log$_{10}$ scale. The parameters used here match those in Figure 11

It is unsurprising that as the ratio $\tilde{c}_{\text{right}}/\tilde{c}_{\text{left}}$ is increased, the power increases. This is expected since there are more available ions for transport. Also, at low concentrations, when $\tilde{i}_{\tilde{V}=0}$ and $\tilde{g}_{\tilde{V}=0}$ are (almost-)constant, the power density is (almost-)constant for all the salts.

The response of KCl is the simplest. At high concentrations, as can be expected, when the $\tilde{i}_{\tilde{V}=0} \sim const$ is saturated (Figure 11), the continued increase of the conductance, $\tilde{g}_{\tilde{V}=0} \sim \tilde{c}_{\text{left}}$, (Figure 9 and Figure 12) results in the decrease of the power $\tilde{P}_{\tilde{V}=0} \sim \tilde{c}^{-1}_{\text{left}}$ (Figure 13). Interestingly, at intermediate concentrations, the power has a maximal value (where the exact value depends on the ratio $\tilde{c}_{right}/\tilde{c}_{left}$) – this is the optimal point for RED using KCl. How to leverage this point should be considered in future works.

Like KCl, HCl doesn't have a crossover point for $\tilde{V}_{\tilde{i}=0}$. Unlike KCl, $\tilde{V}_{\tilde{i}=0}$ doesn't go to zero, resulting in $\tilde{i}_{\tilde{V}=0} \sim \tilde{c}_{left}$ at high concentrations. Here, too, we have $\tilde{g}_{\tilde{V}=0} \sim \tilde{c}_{left}$. However, now, we find that $\tilde{P}_{\tilde{V}=0} \sim \tilde{c}_{left}$. As a result, the power density of HCl, surprisingly, increases at higher concentrations (Figure 13). Perhaps it is important to emphasize that this power density is not due to selectivity, which at this point is virtually zero. Rather, the current is simply a result of the asymmetry in the diffusion coefficients, where the positive ions move more freely than the negative ions and thus induce an electrical current.

The behavior of NaCl and LiCl are the most complicated. Similar to HCl, these salts also have the $\tilde{P}_{\tilde{V}=0} \sim \tilde{c}_{left}$ scaling at large concentrations. However, in contrast to HCl, whose $\tilde{V}_{\tilde{i}=0}$ and $\tilde{i}_{\tilde{V}=0}$ did not exhibit a sign change, NaCl and LiCl, have such a sign change where both $\tilde{V}_{\tilde{i}=0}$ and $\tilde{i}_{\tilde{V}=0}$ change signs. Thus, at the point that $\tilde{i}_{\tilde{V}=0}$ changes the sign such that $\tilde{i}_{\tilde{V}=0}=0$, the power density is zero and this system cannot be used for energy-harvesting purposes.

## VI. CONCLUSIONS

In this work, we have reviewed the classical, paradigmatical GHK theory. Our starting point was to show that GHK's underlying assumption of a uniform electric is inappropriate for most physical systems [Figure 5(b)]. Using non-approximated numerical simulations, we have shown the uniform electric field assumption inadvertently leads to the inconsistent calculation of several quantities (such as the $\tilde{i}-\tilde{V}$ and its derivative $\tilde{g}_{Ohmic}$ and $\tilde{V}_{\tilde{i}=0}$), and depending on the chosen parameters, the errors can be up to 200% (Figure 4), and in some cases, the predicted fluxes [like $j_{GHK}$, Figure 5(c)] look nothing like the numerically calculated solutions of the non-approximated equations. While GHK is a mathematically consistent model, it does not account for the physics correctly. In particular, it is unable to conserve electroneutrality or calculate the electric field in a self-consistent manner.

Independent of the assumption of a uniform electric field or conservation of electroneutrality, GHK was derived with the explicit assumption of a completely charge-less nanopore, $\tilde{\Sigma}_s \triangleq 0$. In most realistic scenarios, $\tilde{\Sigma}_s \neq 0$. Also, until now, it was impossible to connect between GHK and RED (which operates in the opposite limit of $\tilde{\Sigma}_s \gg \tilde{c}_{bulk}$).

To connect GHK and RED, we leveraged our past work [33] in which we derived analytical expressions for $\tilde{g}_{\text{Ohmic}}$, $\tilde{V}_{\tilde{i}=0}$ and $\tilde{i}_{\tilde{V}=0}$ for arbitrary values of $\tilde{\Sigma}_s$ but for symmetric diffusion coefficients (discussed in Sec. V.B.2 and shown in Figure 8 and Figure 9). Here, we have extended this solution for arbitrary values of the diffusion coefficients (Sec. V.A). Admittedly, the mathematical complexity is substantially more difficult than the GHK theory and, as such, might be considered to be less attractive. However, as we have demonstrated throughout this work, the results predicted by this new theory are entirely internally self-consistent and conserve electroneutrality.

The new model can be reduced to three different models [Sec. V.B]. Each model provides remarkable insight into that particular model. However, the insights also add up to provide a very clear physical picture of the general behavior of the electrical conductance. Importantly, we have corroborated our theory using non-approximated numerical simulations and shown remarkable correspondence (Figure 6-Figure 11).

In this work, we have not only analyzed $\tilde{g}_{\tilde{i}=0}$, but we have also analyzed $\tilde{g}_{\tilde{V}=0}$ (Figure 12) and $\tilde{P}_{\tilde{V}=0}$ (Figure 13). While $\tilde{g}_{\tilde{V}=0}$ behaves very similarly to $\tilde{g}_{\tilde{i}=0}$, they are not the same. The subtle difference can lead to substantial errors when calculating the power density. Combined with the behavior of $\tilde{i}_{\tilde{V}=0}$, we have also shown that $\tilde{P}_{\tilde{V}=0}$ also behaves in a rather peculiar manner as a function of the salt currents. In particular, there is a crossover point where the power density is zero.

Undoubtedly, many important challenges related to ion transport remain unresolved. In fact, it might even be more discouraging that one of the greatest challenges, which was supposedly resolved eight decades ago, has been reopened. However, this work provides the essential framework needed to reanalyze all experimental and numerical data that was previously analyzed using GHK.

Furthermore, this work can be used to address several new/old questions. First, since GHK makes incorrect predictions for two species, there is no doubt it will also make incorrect predictions regarding multiple species. Thus, future works should utilize the mathematical method proposed here to derive a multiple-species solution. Second, in the last decade, many new 2D and atomistically smooth materials have been introduced. The surface charge of these materials is regulated by the material properties and the electrolyte properties [4,47,52,54–58]. This work should be extended to account for the nonlinear change and perhaps spatial dependency of $\tilde{\Sigma}_s(\mathbf{x})$ on all the various parameters. Finally, in this work, we have, as GHK did in the past, ignored the effects of convection. There is no doubt that advection will vary $\tilde{i} - \tilde{V}$ in a vary non-trivial manner. Future works should also account for advection.

We conclude with a quote attributed to Sir Lawerence Bragg [59]

> *"The fun in science lies not in discovering facts, but in discovering new ways of thinking about them. The test which we apply to these ideas is this — do they enable us to fit the facts to each other, and see that more and more of them can be explained by fewer and fewer fundamental laws."*

Here, we have shown that old ideas can and should be revisited. These lead to new ideas that better fit the facts while utilizing a smaller set of fundamental concepts. We hope that we have been able to show that there are many new ways to think about ion transport through nanoporous medium.

**Acknowledgments.** We thank Prof. Slaven Garaj for suggesting that we revisit the GHK problem and Mr. Oren Lavi for enlightening discussions. We are extremely grateful to Prof. Michael Mond and Prof. James Butles for their careful reading of the early versions of this work and the ensuing discussions. In particular, we are indebted to Prof. Mond for several clarifying comments on Goldman's work. This work was supported by Israel Science Foundation grants 337/20 and 1953/20. We acknowledge the support of the Ilse Katz Institute for Nanoscale Science and Technology and the Pearlstone Center for Aeronautical Engineering Studies.

### APPENDIX A: NUMERICAL SIMULATIONS

We solved Eqs. (7)-(9) for arbitrary salt gradients $(\tilde{c}_{\text{left}}, \tilde{c}_{\text{right}})$ and diffusion coefficients $(\tilde{D}_+, \tilde{D}_-)$ under a potential drop, $\tilde{V}$, using the Comsol modules Transport of Diluted Species and Electrostatics in a simple 1D setup. A detailed description of the numerical method is given in our previous work [33], where the sole difference between this work and that work is that there $\tilde{D}_+ \equiv \tilde{D}_-$ and here, this is not the default case. Except for $\tilde{V}$, the only other parameters that were varied were $\tilde{c}_{\text{left}}$, $\tilde{c}_{\text{right}}$ and $\tilde{D}_+$. Our default diffusion coefficient for the negative species was that of Cl, $\tilde{D}_- = \tilde{D}_{\text{Cl}}$. Table 2 gives the default values for the simulations, while the figure captions and figure legends, throughout this work, give the values of $\tilde{V}$, $\tilde{c}_{\text{left}}$, $\tilde{c}_{\text{right}}$ and $\tilde{D}_-$.

Table 2: Simulation parameters. Diffusion coefficients were taken from Ref. [60].

| Parameter | Value | Units |
|---|---|---|
| Temperature, $\tilde{T}$ | 298 | [K] |
| Channel length, $\tilde{L}$ | $10^{-3}$ | [m] |
| Excess counterion concentration, $\tilde{\Sigma}_s$ | 1 | [mol/m$^3$] |
| Relative permittivity, $\varepsilon_r$ | 80 | |
| Valency, $z$ | 1 | |
| Cl$^-$ diffusion coefficient, $\tilde{D}_- = \tilde{D}_{Cl}$ | $2.03 \cdot 10^{-9}$ | [m$^2$/s] |
| Li$^+$ diffusion coefficient, $\tilde{D}_{Li}$ | $1.029 \cdot 10^{-9}$ | [m$^2$/s] |
| K$^+$ diffusion coefficient[1], $\tilde{D}_K$ | $2.03 \cdot 10^{-9}$ | [m$^2$/s] |
| Na$^+$ diffusion coefficient, $\tilde{D}_{Na}$ | $1.33 \cdot 10^{-9}$ | [m$^2$/s] |
| H$^+$ diffusion coefficient, $\tilde{D}_H$ | $9.31 \cdot 10^{-9}$ | [m$^2$/s] |

[1] In reality, $\tilde{D}_K = 1.96 \cdot 10^{-9}$. However, for a purely symmetric salt ($\tilde{D}_+ \equiv \tilde{D}_-$), we insert into $\tilde{D}_K$ the value of $\tilde{D}_{Cl}$.